\documentclass[reprint,amsmath,amssymb,aps,nofootinbib,]{revtex4-2}
\usepackage{graphicx}
\usepackage{dcolumn}
\usepackage{bm}
\usepackage{amsmath}
\usepackage{mathrsfs}
\usepackage{xcolor} 
\usepackage{slashed}  
\usepackage{multirow}
\usepackage{booktabs}
\usepackage[export]{adjustbox}         
\usepackage[
    colorlinks=true,
    linkcolor=blue,
    citecolor=blue,
    urlcolor=blue
]{hyperref}
\newcommand{\orcid}[1]{\hspace{1mm}\href{https://orcid.org/#1}{\includegraphics[height=0.3cm,keepaspectratio]{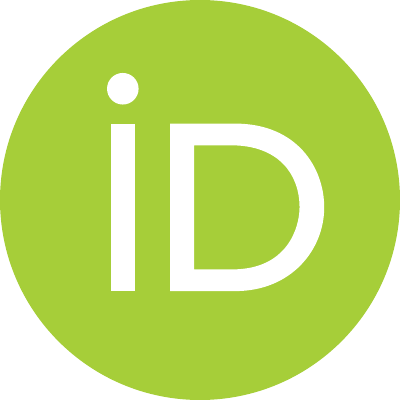}}}

\begin{document}
\hfill
\makebox[0pt][r]{IFJPAN-IV-2026-15}
\preprint{IFJPAN-IV-2026-15}
\vspace{6pt}

\title{Can $\gamma\gamma$ collisions rival $e^-e^+$ in probing doubly charged Higgs bosons?}

\author{Abdesslam Arhrib\orcid{0000-0001-5619-7189}}
\email{aarhrib@gmail.com}
\affiliation{Abdelmalek Essaadi University, FST Tanger B.P. 416, Morocco}
\affiliation{LAPTh, CNRS, Universit\'e Savoie Mont-Blanc, 9 Chemin de Bellevue, 74940, Annecy, France.}

\author{Rachid Benbrik\orcid{0000-0002-5159-0325}}
\email{r.benbrik@uca.ac.ma}
\affiliation{Laboratory of Physics, Energy, Environment, and Applications, Cadi Ayyad University, Sidi Bouzid, P.O. Box 4162, Safi, Morocco}

\author{Mohammed Boukidi\orcid{0000-0001-9961-8772}}
\email{mohammed.boukidi@ifj.edu.pl}
\affiliation{Institute of Nuclear Physics, Polish Academy of Sciences, ul. Radzikowskiego 152, Cracow, 31-342, Poland}

\author{Mohamed Chabab\orcid{0000-0002-2772-4290}}
\email{mchabab@uca.ac.ma}
\affiliation{Cadi Ayyad University,  Faculty of Science Semlalia, LPHEAG, P.O.B. 2390 Marrakech 40000, Morocco.}
\affiliation{Cadi Ayyad University, National School of Applied Science, P.O.B. 63 Safi 46000, Morocco}
\author{Khalid Goure\orcid{0009-0007-5292-5012}}
\email{khalidgoure01@gmail.com}
\affiliation{Cadi Ayyad University,  Faculty of Science Semlalia, LPHEAG, P.O.B. 2390 Marrakech 40000, Morocco}

\author{Stefano Moretti\orcid{0000-0002-8601-7246}}
\email{stefano.moretti@physics.uu.se/s.moretti@soton.ac.uk}
\affiliation{Department of Physics and Astronomy, Uppsala University, Box 516, SE-751 20 Uppsala, Sweden}
\affiliation{School of Physics and Astronomy, University of Southampton, Southampton, SO17 1BJ, United Kingdom}
\date{\today}

\begin{abstract}
High-energy $\gamma\gamma$ collisions, realizable as an operational mode of future lepton linear colliders such as the ILC and CLIC, provide a promising environment to probe extended Higgs sectors. We investigate the sensitivity of such colliders to doubly charged Higgs bosons within the 2-Higgs Doublet Model with type-II seesaw (2HDMcT). Focusing on the three-body production channels $\gamma\gamma \to H^{\pm\pm}H_1^{\mp}H_1^{\mp}$ and $\gamma\gamma \to H^{\pm\pm}H_1^{\mp}W^{\mp}$, we perform a parameter space scan consistent with theoretical constraints as well as current collider, flavor, and electroweak precision observables~(EWPOs). We show that $\gamma\gamma$ collisions can rival the discovery potential of the conventional $e^+e^-$ mode for probing doubly charged Higgs bosons through a $4\ell+\slashed{E}_T$ signature~($\ell=e,\mu$). Despite the reduced effective luminosity resulting from the photon spectrum, the significantly enhanced production cross sections, exceeding those in electron-positron collisions by more than one order of magnitude, compensate for this limitation. By performing a detailed signal-to-background analysis at center-of-mass energies of $\sqrt{s}=830$ and $1245$ GeV, we demonstrate that a discovery significance at the $5\sigma$ level can be achieved for viable benchmark points~(BPs). 
\end{abstract}

\maketitle

\section{Introduction}

The discovery of a Higgs boson with a mass of approximately $125~\mathrm{GeV}$ by the ATLAS and CMS collaborations marked a major success of the Standard Model~(SM) and completed its particle spectrum. However, despite this achievement, the SM still leaves several fundamental questions unanswered. Among these are the origin of neutrino masses, the nature of dark matter, and the mechanism responsible for the observed matter--antimatter asymmetry in the Universe~\cite{Zwicky:1933gu,Rubin:1970zza,Veltman:1980mj,SupernovaSearchTeam:1998fmf,Crivellin:2023zui}. These open issues strongly motivate the search for physics beyond the SM~(BSM). One promising direction is provided by extended Higgs-sector frameworks containing additional scalar multiplets.

Seesaw mechanisms provide an appealing explanation for the origin of neutrino masses. Among the BSM scenarios that realize such mechanisms, the Two-Higgs-Doublet Model with a Complex Triplet (2HDMcT) is particularly attractive: the SM scalar sector is extended by a second Higgs doublet and a complex scalar triplet. Besides enriching the Higgs spectrum, this scenario naturally accommodates neutrino masses through the type-II seesaw mechanism. The model predicts additional neutral, singly charged, and doubly charged Higgs bosons. The charged states are particularly distinctive, since their observation would provide direct evidence for electroweak scalar representations beyond the SM doublet structure. Consequently, searches for charged Higgs bosons, especially doubly charged states, form an important part of the physics program at present and future colliders.

While hadron colliders generally benefit from high center-of-mass energies, their complex QCD environment can limit the sensitivity to challenging signatures. In contrast, future lepton colliders offer a much cleaner experimental environment, allowing for more precise measurements and improved background control. For this reason, they are regarded as ideal machines for investigating the detailed properties of non-standard Higgs bosons. 

In the past few years, considerable attention has been devoted to the study of doubly charged Higgs boson production and decay at both hadron and lepton colliders\footnote{Here, lepton colliders include both $e^-e^+$ and $\gamma\gamma$ collision modes.}~\cite{Melfo:2011nx,Huitu:1996su,Gunion:1996pq,Chakrabarti:1998qy,Muhlleitner:2003me,Chun:2003ej,Akeroyd:2005gt,Han:2007bk,Akeroyd:2007zv,delAguila:2008cj,FileviezPerez:2008jbu,CiezaMontalvo:2008ew,Akeroyd:2009hb,Akeroyd:2010ip,Aoki:2011pz,Arhrib:2011uy,Akeroyd:2011zza,Chiang:2012dk,Sugiyama:2012yw,Akeroyd:2012nd,Chun:2012zu,delAguila:2013mia,Chun:2013vma,Kanemura:2013vxa,Kanemura:2014goa,Kanemura:2014ipa,Dutta:2014dba,Kang:2014lwn,Han:2015hba,Han:2015sca,Mitra:2016wpr,Ghosh:2017pxl,Antusch:2018svb,BhupalDev:2018tox,deMelo:2019asm,Primulando:2019evb,Chun:2019hce,Fuks:2019clu,Padhan:2019jlc,Ashanujjaman:2021txz,Ashanujjaman:2022ofg,Cakir:2006pa,Nomura:2017abh,Blunier:2016peh,Crivellin:2018ahj,Agrawal:2018pci,Ashanujjaman:2022tdn,Ruiz:2022sct,Chiang:2025lab,Dev:2019hev,Yang:2021skb,Deppisch:2015qwa,Das:2023tna,Li:2023ksw,Maharathy:2023dtp,Jueid:2023qcf,Belfkir:2023lot,Jia:2024wqi,	CMS:2012dun,ATLAS:2012hi,CMS:2014mra,BhupalDev:2013xol,ATLAS:2014kca,ATLAS:2014vih,ATLAS:2017xqs,CMS:2017fhs,ATLAS:2021jol,ATLAS:2023sua,ATLAS:2024txt,Guedes:2025jqu,Englert:2026eou,Das:2026qwe,Li:2026znl,BrahimAit-Ouazghour:2026stq,Zhou:2026hdt}. 

In addition to the conventional $e^-e^+$ mode, future linear colliders can operate in a high-energy $\gamma\gamma$ mode through Compton backscattering of intense laser light off energetic electron beams. This provides a complementary avenue for exploring BSM physics. The $\gamma\gamma$ mode is particularly well suited for studying charged particles because photons couple directly to electrically charged states. As a consequence, charged Higgs bosons can be produced efficiently through electromagnetic interactions.

The $\gamma\gamma$ mode is particularly important for doubly charged Higgs bosons because their interaction strength increases with the square of the electric charge, so that the production rate of $H^{\pm\pm}$ states receives a substantial enhancement compared to the corresponding singly charged particles $H^\pm$. This makes photon colliders exceptionally sensitive to triplet-like scalar states predicted in models such as the 2HDMcT. Furthermore, the cleaner environment of $\gamma\gamma$ collisions, together with the possibility of polarized photon beams, provides powerful tools for suppressing SM backgrounds and improving signal extraction.

The effective center-of-mass energy of a photon collider can reach nearly $83\%$ of the parent $e^-e^+$ beam energy, making it possible to probe heavy charged Higgs bosons in the TeV range\footnote{Accordingly, in this work we evaluate the production cross section at this peak energy.}. These advantages have motivated extensive investigations of charged Higgs production in photon collisions within various BSM scenarios.

In this work we focus on the 2HDMcT~\cite{Chen:2014xva,Ouazghour:2018mld,Ouazghour:2023eqr,Ouazghour:2024fgo,AitOuazghour:2025cmd,BrahimAit-Ouazghour:2024dpr,BrahimAit-Ouazghour:2025xap}. Compared to the 2HDM scalar sector, the 2HDMcT features a broader spectrum and notable signatures, with the doubly charged Higgs $H^{\pm\pm}$ providing a particularly clean ``smoking-gun'' signature. Furthermore, the 2HDMcT can be viewed as one of the simplest and most economical extensions of the SM capable of accommodating neutrino masses~\cite{FileviezPerez:2008jbu,King:2015aea} and addressing the DM problem~\cite{Chen:2014lla}. In addition, the 2HDMcT can be distinguished from the minimal Higgs-Triplet Model (HTM) via certain cascade topologies.  In the 2HDMcT the chain\footnote{Hereafter, the subscripts 1 and 2 refer to singly charged Higgs mass eigenstates with $m_{H_2^\pm}>m_{H_1^\pm}$.} $pp \to Z/\gamma\to H_2^{+}H_2^{-}\to H^{++}W^{-}\,H^{--}W^{+}\to\ell^{+}\ell^{+}\ell^{-}\ell^{-}+4j$ can proceed fully on shell. This is because the additional scalars relax the oblique-parameter bounds and allow mass splittings $\Delta m = m_{H^\pm} - m_{H^{\pm\pm}}$ of order ${\cal O}(m_W)$ for $H_2^\pm$. In the HTM, EWPOs constrain $|\Delta m|\lesssim40$ GeV\,\cite{Melfo:2011nx,Arhrib:2011uy,Ashanujjaman:2021txz,Ashanujjaman:2025scr}, forcing the $W$ bosons to be off-shell~\cite{Ashanujjaman:2025scr} in the analogous sequence $pp\to Z/\gamma\to H^{+}H^{-}\to H^{++}W^{-*}\,H^{--}W^{+*}$ and severely suppressing its rate. The large mass splitting in the 2HDMcT allows  the on-shell decay channels $H^{\pm\pm}\to W^{\pm}H_1^{\pm}$ and $H^{\pm\pm}\to H_1^{\pm}H_1^{\pm}$, thereby making the associated production modes $H^{\pm\pm}H_1^{\mp}H_1^{\mp}$ and $H^{\pm\pm}H_1^{\mp}W^{\mp}$ promising discovery channels for $H^{\pm\pm}$ at current and future colliders. These processes have already been shown to provide promising signatures at $e^-e^+$ colliders~(see Ref.~\cite{BrahimAit-Ouazghour:2026stq}). Consequently, the corresponding three-body associated production channels in $\gamma\gamma$ collisions are also expected to play an important role in probing the $H^{\pm\pm}$ states of the 2HDMcT. It has also been shown that interactions between the doublet and triplet fields can induce a Strong First-Order EW Phase Transition (SFOEWPT), thereby enabling EW baryogenesis~\cite{Ramsey-Musolf:2019lsf}.
In addition, the presence of the second Higgs doublet leads to significant changes in the production and decay properties of $H^{\pm\pm}$ states  in the 2HDMcT compared to the HTM (see Ref.~\cite{Chen:2014qda}).

In this study, we investigate the sensitivity of $\gamma\gamma$ colliders to doubly charged Higgs bosons within the 2HDMcT. We focus on the three-body production channels
$\gamma\gamma \to H^{\pm\pm}H_1^{\mp}H_1^{\mp}$ and
$\gamma\gamma \to H^{\pm\pm}H_1^{\mp}W^{\mp}$,
performing a parameter space scan consistent with theoretical constraints, as well as current collider, flavor, and EWPOs. A full Monte Carlo (MC) simulation at the detector level is carried out to assess the sensitivity of these channels at a future $\gamma\gamma$ collider using an ILC detector setup~\cite{ILCInternationalDevelopmentTeam:2022izu}\footnote{We use the ILC detector card because a dedicated implementation is available; the same analysis strategy can be adapted to other TeV-scale linear-collider setups, such as CLIC~\cite{Linssen:2012hp}.}, considering center-of-mass energies of $830$ and $1245~\text{GeV}$.

This paper is organized as follows. In Sect.~\ref{gaga}, we introduce the $\gamma\gamma$ collider setup. In Sect.~\ref{prese_2HDMcT}, we briefly review the 2HDMcT model. Sect.~\ref{constraint} summarizes the theoretical and experimental constraints imposed on the parameter space. Our computational procedure and numerical results are presented in Sects.~\ref{COMPUTATIONAL_PROCEDURE} and \ref{Sec:Results}, respectively. Finally, Sect.~\ref{conlusion} is devoted to our conclusions.

\section{$\gamma \gamma$ colliders}
\label{gaga}
The concept of a high-energy $\gamma\gamma$ collider based on an $e^{-}e^{+}$  linear machine was originally proposed in the early 1980s~\cite{Ginzburg:1981ik,Ginzburg:1981vm,Ginzburg:1982yr,Telnov:1989sd}. The basic idea relies on generating energetic photon beams through Compton backscattering of laser photons off high-energy electron (or positron) beams before the interaction point\footnote{For comparison with the $e^+e^-$ mode, the electron and positron beams are assumed to be unpolarized. Polarization effects, which can enhance the sensitivity of future $e^+e^-$ linear colliders, are not considered in this analysis.}. This technique has since been extensively investigated in the context of such machines~\cite{Telnov:1995hc,Telnov:1997vg,Telnov:1998vs,Ginzburg:1999wz,Telnov:2000kq,Telnov:2000zx,Telnov:2000ep,Asner:2001vh,ECFADESYPhotonColliderWorkingGroup:2001ikq,Burkhardt:2002vh,CLICPhysicsWorkingGroup:2004qvu,Telnov:2006cj,Telnov:2008zz,Telnov:2009vq,Yu:2018omk}.

In the conversion region, a laser photon with energy $E_{0}$ collides almost head-on with an electron beam of energy $E_{e}$. The scattered photons acquire a large fraction of the electron-beam energy, with the maximum photon energy given by

\begin{equation}
E^{\mathrm{max}}_{\gamma}=
\frac{\kappa}{1+\kappa}E_{e},
\qquad
\kappa=
\frac{4E_{e}E_{0}\cos^{2}(\alpha/2)}{m_{e}^{2}},
\label{eq:max_photon_energy}
\end{equation}

where $m_e$ denotes the electron mass and $\alpha$ is the collision angle. The optimal configuration corresponds to the pair-production threshold, leading to $\kappa^{\rm max}\simeq4.83$ and consequently

\begin{equation}\label{eq:Emax}
E_{\gamma}^{\mathrm{max}}\simeq0.83\,E_e.
\end{equation}

In the present analysis, we do not perform the full convolution of the subprocess cross section with the Compton backscattered photon spectra. Instead, the cross section is evaluated at the peak of the photon spectrum, corresponding to Eq.~(\ref{eq:Emax}), and the event rate is estimated by assuming an effective high-energy $\gamma\gamma$ luminosity equal to approximately $10\%$ of the parent $e^-e^+$ luminosity as follows\footnote{Here, $10\%$ is a conservative estimate: depending on the $\gamma\gamma$ collider setup the luminosity in the high energy part of the spectrum can reach in some configurations $(1/3)L_{e^+e^-}$ (see Ref.~\cite{Telnov:2000zx}).}:
\begin{equation}
\mathcal{L}_{\gamma\gamma}(y>0.8\,y^{\rm max})
\simeq0.1\,\mathcal{L}_{ee},
\end{equation}
where $y=E_{\gamma\gamma}/(2E_e)$ and $y^{\rm max}\simeq0.83$.	
This choice is adopted as a conservative phenomenological configuration, providing a simple estimate of the production rate in the high-energy region relevant for new-physics searches while avoiding the complexity of a full spectral convolution.

In short, in the framework of the 2HDMcT, $\gamma\gamma$ collisions provide a particularly clean environment for investigating heavy scalar states, such as the doubly charged Higgs boson $H^{++}$, especially in parameter regions that remain challenging at hadron colliders. While accurate predictions require folding the cross sections with the photon luminosity spectrum, phenomenological studies often employ the peak-energy approximation\footnote{See, for example, Ref.~\cite{Garcia-Abenza:2020xkk}}, which captures the dominant features of the process, as we do here.

\vspace*{6pt}
\section{2HDMcT: BRIEF REVIEW}

\label{prese_2HDMcT}

The 2HDMcT includes two Higgs doublets $\Phi_{i}$ ($i = 1,2$)	and one colorless scalar field $\Delta$ transforming as a triplet under the $SU(2)_L$ gauge group with hypercharge $Y_\Delta=2$. The most general gauge-invariant Lagrangian of the 2HDMcT is given by 

\begin{equation}
\begin{matrix}
\mathcal{L}=\sum_{i=1}^2(D_\mu{\Phi_i})^\dagger(D^\mu{\Phi_i})+Tr(D_\mu{\Delta})^\dagger(D^\mu{\Delta})\vspace*{0.12cm}\\
\hspace{1cm}-V(\Phi_i, \Delta)+\mathcal{L}_{\rm Yukawa}.
\label{eq:thdmt-lag}
\end{matrix}
\end{equation}
\vspace*{10pt}

The covariant derivatives are defined as
\begin{equation}
D_\mu{\Phi_i}=\partial_\mu{\Phi_i}+igT^a{W}^a_\mu{\Phi_i}+i\frac{g'}{2}B_\mu{\Phi_i}, \label{eq:covd1}
\end{equation}
\vspace*{-1cm}
\begin{equation}
~~D_\mu{\Delta}=\partial_\mu{\Delta}+ig[T^a{W}^a_\mu,\Delta]+ig' \frac{Y_\Delta}{2} B_\mu{\Delta}, \label{eq:covd2}
\end{equation} 
with (${W}^a_\mu$, $g$) and ($B_\mu$, $g'$) denoting the gauge fields and couplings of $SU(2)_L$ and $U(1)_Y$, respectively, while $T^a \equiv \sigma^a/2$, with $\sigma^a$ ($a=1,2,3$) denoting the Pauli matrices. The 2HDMcT Higgs potential is given by \cite{Chen:2014xva,Ouazghour:2018mld}:
\begin{widetext}
	\begin{equation}
	\begin{aligned}
	&V(\Phi_1,\Phi_2,\Delta) = m_{11}^2 \Phi_1^\dagger\Phi_1
	+ m_{22}^2\Phi_2^\dagger\Phi_2
	- \left[m_{12}^2\Phi_1^\dagger\Phi_2 + {h.c.}\right]
	+ \frac{\lambda_1}{2}(\Phi_1^\dagger\Phi_1)^2
	+ \frac{\lambda_2}{2}(\Phi_2^\dagger\Phi_2)^2
	+\lambda_3(\Phi_1^\dagger\Phi_1)(\Phi_2^\dagger\Phi_2)+\lambda_4(\Phi_1^\dagger\Phi_2)(\Phi_2^\dagger\Phi_1)
	\\
	&~
	+ \left\{
	\frac{\lambda_5}{2}(\Phi_1^\dagger\Phi_2)^2
	+ \big[\beta_1(\Phi_1^\dagger\Phi_1)
	+\beta_2(\Phi_2^\dagger\Phi_2)\big]\Phi_1^\dagger\Phi_2+ {h.c.}\right\} + \lambda_6\,\Phi_1^\dagger \Phi_1 \, \mathrm{Tr}(\Delta^{\dagger}\Delta)
	+ \lambda_7\,\Phi_2^\dagger \Phi_2 \, \mathrm{Tr}(\Delta^{\dagger}\Delta)
	+ \lambda_8\,\Phi_1^\dagger \Delta \Delta^{\dagger} \Phi_1
	\\
	&~
	+ \lambda_9\,\Phi_2^\dagger \Delta \Delta^{\dagger} \Phi_2
	+ m^2_{\Delta}\, \mathrm{Tr}(\Delta^{\dagger}\Delta)
	+ \bar{\lambda}_8 \left(\mathrm{Tr}\,\Delta^{\dagger}\Delta\right)^2
	+ \bar{\lambda}_9 \, \mathrm{Tr}\left(\Delta^{\dagger}\Delta\right)^2
	+ \left\{\mu_1 \Phi_1^T i\sigma^2 \Delta^{\dagger}\Phi_1 + \mu_2 \Phi_2^T i\sigma^2 \Delta^{\dagger}\Phi_2 + \mu_3 \Phi_1^T i\sigma^2 \Delta^{\dagger}\Phi_2 + {h.c.}\right\},
	\end{aligned}
	\label{scalar_pot}
	\end{equation}
\end{widetext}

where $Tr$ denotes the trace over $2\times2$ matrices. The Higgs doublets $\Phi_{i}$ and triplet $\Delta$ are represented by 
\begin{equation}
\vspace*{-0.5truecm}
\Delta=\left(
\begin{array}{cc}
\delta^+/\sqrt{2} & \delta^{++} \\
(v_t+\delta^0+i\eta_0)/\sqrt{2} & -\delta^+/\sqrt{2}\\
\end{array}
\right),\end{equation}
\vspace*{-0.25truecm}
\begin{equation}
\Phi_1=\left(
\begin{array}{c}
\phi_1^+ \\
\phi^0_1 \\
\end{array}
\right){,}~~~\Phi_2=\left(
\begin{array}{c}
\phi_2^+ \\
\phi^0_2 \\
\end{array}
\right),\end{equation}

with $\phi^0_1=(v_1+\rho_1+i\eta_1)/\sqrt{2}$ and $\phi^0_2=(v_2+\rho_2+i\eta_2)/\sqrt{2}$. Following spontaneous electroweak symmetry breaking (EWSB), the two doublets and the triplet acquire vacuum expectation values (VEVs) $v_1$, $v_2$, and $v_t$, respectively. The physical scalar spectrum then contains three CP-even neutral Higgs bosons $(h_1,h_2,h_3)$, two pairs of singly charged Higgs bosons~\cite{BrahimAit-Ouazghour:2025xap} $(H_1^{\pm},H_2^{\pm})$, two CP-odd neutral Higgs bosons $(A_1,A_2)$, and one pair of doubly charged Higgs bosons $H^{\pm\pm}$~\cite{BrahimAit-Ouazghour:2026stq}.  

Expanding the covariant derivative ${D}_\mu$ and expressing the gauge and scalar fields in terms of the physical mass eigenstates, the couplings of the neutral Higgs bosons $h_i$ ($i=1,2,3$) to the SM massive gauge bosons $V=W,Z$ can be extracted, as summarized in Tab.~\ref{table2}. Unlike the 2HDM, the triplet field $\Delta$ couples directly to the SM gauge bosons, leading to an additional contribution. As a result, the couplings $C^{h_i}_V$ ($V=W^\pm,Z$) differ by a factor of 2 in the terms proportional to $v_t$.
\vspace*{-6pt}
\subsection{The Yukawa sector}
The Yukawa Lagrangian of the 2HDMcT contains the complete Yukawa sector of the 2HDM along with an additional term $\mathcal{L}_{\rm Triplet}$ given by
\begin{eqnarray}
	-\mathcal{L}_{\rm Triplet} = Y_{ij}L_i^TCi\sigma_2\Delta L_j + h.c.,
	\label{neutrino_mass}
\end{eqnarray}
where $Y$ is a $3\times3$ complex symmetric matrix and $L = (\nu_L,l_L)^T$ denotes the $SU(2)_L$ doublets of left-handed leptons.
This term generates tiny Majorana masses for the neutrinos after EWSB as follows:
\vspace*{-6pt}
\begin{eqnarray}
m_\nu = \sqrt{2}\,Y v_\Delta.
\end{eqnarray}
\vspace*{-3pt}
The $3 \times 3$ neutrino mass matrix, $m_\nu$, can be diagonalized through the unitary 
transformation involving the Pontecorvo--Maki--Nakagawa--Sakata (PMNS) matrix, 
$U_{\mathrm{PMNS}}$, according to
\begin{equation}
U_{\mathrm{PMNS}}^T \, m_\nu \, U_{\mathrm{PMNS}} 
= m_\nu^d 
= \mathrm{diag}(m_1, m_2, m_3) \, ,
\end{equation}
where $m_1, m_2$, and $m_3$ denote the three neutrino mass eigenvalues. 
The matrix $U_{\mathrm{PMNS}}$ is conventionally parameterized by
three mixing angles ($\theta_{12}, \theta_{23}, \theta_{13}$), 
one Dirac CP-violating phase ($\delta$), 
and two Majorana phases ($\phi_1, \phi_2$) as

\small	
\begin{widetext}
	\begin{equation}
	U_{\rm PMNS} =
	\begin{pmatrix}
	c_{12} c_{13} & s_{12} c_{13} & s_{13} e^{-i\delta} \\[6pt]
	- c_{12} s_{13} s_{23} e^{i\delta} - c_{23} s_{12} & 
	c_{12} c_{23} - s_{12} s_{13} s_{23} e^{i\delta} & 
	c_{13} s_{23} \\[6pt]
	s_{12} s_{23} - c_{12} c_{23} s_{13} e^{i\delta} & 
	- c_{23} s_{12} s_{13} e^{i\delta} - c_{12} s_{23} & 
	c_{13} c_{23}
	\end{pmatrix}
	\cdot 
	\mathrm{diag}\!\left(e^{i\Phi_1/2},\, 1,\, e^{i\Phi_2/2}\right).
	\end{equation}
\end{widetext}
\normalsize
In the normal hierarchy~(NH), $m_{\nu_1} < m_{\nu_2} < m_{\nu_3}$. The free parameters of the neutrino sector can be expressed in terms of the neutrino oscillation parameters together with the mass of the lightest neutrino:
\begin{equation}
m_{\nu_1},  \ \theta_{12}, \ \theta_{13}, \ \theta_{23},\ \Delta m_{21}^2, \ \Delta m_{31}^2, \ \delta, \ \phi_1, \ \phi_2.
\end{equation}
Here, $\Delta m_{31}^2>0$ and $\Delta m_{21}^2>0$.

The parameters $\Delta m_{21}^2$, $\Delta m_{31}^2$, $\theta_{12}$, $\theta_{13}$, $\theta_{23}$, and $\delta$ are largely constrained by global fits to oscillation data~\cite{Esteban:2024eli}, 
while the lightest neutrino mass $m_{\nu_{\min}}=m_{\nu_1}$ and the two Majorana phases $\phi_1, \phi_2$ are not constrained by current oscillation data. 

The heavier neutrino masses are then determined as follows:
\begin{equation}
m_{\nu_2} = \sqrt{m_{\nu_1}^2 + \Delta m_{21}^2}, \qquad 
m_{\nu_3} = \sqrt{m_{\nu_1}^2 + \Delta m_{31}^2}.
\end{equation}

\begin{table*}[t]
	\centering
	\caption{Normalized couplings of the neutral CP-even Higgs bosons $h_i$ to the gauge bosons $V=W,Z$ in the 2HDMcT.}
	\label{table2}
	\renewcommand{\arraystretch}{1.4}
	\begin{tabular}{c c c}
		\toprule
		& $C_W^{h_i}$ & $C_Z^{h_i}$ \\
		\midrule
		$h_1$ &
		$\displaystyle
		\frac{v_1}{v}\mathcal{E}_{11}
		+\frac{v_2}{v}\mathcal{E}_{21}
		+2\frac{v_t}{v}\mathcal{E}_{31}$ &
		$\displaystyle
		\frac{v_1}{v}\mathcal{E}_{11}
		+\frac{v_2}{v}\mathcal{E}_{21}
		+4\frac{v_t}{v}\mathcal{E}_{31}$ \\[2mm]
		
		$h_2$ &
		$\displaystyle
		\frac{v_1}{v}\mathcal{E}_{12}
		+\frac{v_2}{v}\mathcal{E}_{22}
		+2\frac{v_t}{v}\mathcal{E}_{32}$ &
		$\displaystyle
		\frac{v_1}{v}\mathcal{E}_{12}
		+\frac{v_2}{v}\mathcal{E}_{22}
		+4\frac{v_t}{v}\mathcal{E}_{32}$ \\[2mm]
		
		$h_3$ &
		$\displaystyle
		\frac{v_1}{v}\mathcal{E}_{13}
		+\frac{v_2}{v}\mathcal{E}_{23}
		+2\frac{v_t}{v}\mathcal{E}_{33}$ &
		$\displaystyle
		\frac{v_1}{v}\mathcal{E}_{13}
		+\frac{v_2}{v}\mathcal{E}_{23}
		+4\frac{v_t}{v}\mathcal{E}_{33}$ \\
		\bottomrule
	\end{tabular}
\end{table*}

The doubly charged mass eigenvalue, $m_{H^{\pm\pm}}^2$, corresponding to the doubly charged field eigenstate, $H^{\pm\pm}$, can be determined by collecting all the coefficients of $\delta^{++}\delta^{--}$ in the Higgs potential. This gives
	\begin{align}
	m_{H^{\pm\pm}}^2=&\frac{\sqrt{2}\mu_1 v_1^2 + \sqrt{2}\mu_3 v_1 v_2 + \sqrt{2}\mu_2 v_2^2 - \lambda_8 v_1^2 v_t
		}{2v_t}\nonumber \\
	&+ \frac{- \lambda_9 v_2^2 v_t - 2 \bar{\lambda}_9 v_t^3}{2v_t}.  \label{eq:mHpmpm}
	\end{align}
\begin{table}[h]
	\centering
	\caption{Normalized Yukawa coupling coefficients of the neutral Higgs bosons ($h_i$ and $A_j$) to up-type quarks ($U$), down-type quarks ($D$), and charged leptons ($\ell$) in the Type-X 2HDMcT. The expressions for $\mathcal{E}_{ij}$ and $\mathcal{O}_{ij}$ are given in Ref.~\cite{Ouazghour:2018mld}.}
	\label{table1}
	\renewcommand{\arraystretch}{1.1}
	\begin{tabular}{c c c c c}
		\toprule
		& & $U$ & $D$ & $\ell$ \\
		\midrule
		
		\multirow{3}{*}{$h_i$}
		& $h_1$ &
		$\displaystyle\frac{\mathcal{E}_{12}}{s_\beta}$ &
		$\displaystyle\frac{\mathcal{E}_{12}}{s_\beta}$ &
		$\displaystyle\frac{\mathcal{E}_{11}}{c_\beta}$ \\
		
		& $h_2$ &
		$\displaystyle\frac{\mathcal{E}_{22}}{s_\beta}$ &
		$\displaystyle\frac{\mathcal{E}_{22}}{s_\beta}$ &
		$\displaystyle\frac{\mathcal{E}_{21}}{c_\beta}$ \\
		
		& $h_3$ &
		$\displaystyle\frac{\mathcal{E}_{32}}{s_\beta}$ &
		$\displaystyle\frac{\mathcal{E}_{32}}{s_\beta}$ &
		$\displaystyle\frac{\mathcal{E}_{31}}{c_\beta}$ \\
		
		\midrule
		
		\multirow{2}{*}{$A_j$}
		& $A_1$ &
		$\displaystyle\frac{\mathcal{O}_{22}}{s_\beta}$ &
		$\displaystyle\frac{\mathcal{O}_{22}}{s_\beta}$ &
		$\displaystyle\frac{\mathcal{O}_{21}}{c_\beta}$ \\
		
		& $A_2$ &
		$\displaystyle\frac{\mathcal{O}_{32}}{s_\beta}$ &
		$\displaystyle\frac{\mathcal{O}_{32}}{s_\beta}$ &
		$\displaystyle\frac{\mathcal{O}_{31}}{c_\beta}$ \\
		
		\bottomrule
	\end{tabular}
\end{table}

\vspace{6pt}
\section{Constraints}
\vspace{6pt}
\label{constraint}
The parameter space of the 2HDMcT is subject to the theoretical constraints of Refs.~\cite{Ouazghour:2018mld, Ouazghour:2023eqr}, together with Higgs measurements and collider exclusion limits, as summarized below.
\begin{itemize}
	\item \textbf{Unitarity}: The Higgs and Goldstone $2\to2$ scattering matrix elements must obey conservation of probability.
	\item \textbf{Perturbativity}: The quartic couplings of the Higgs potential are constrained by the following conditions: $| \lambda_i|<8 \pi$.
	\item \textbf{Vacuum Stability}: Boundedness from below (BFB) is required along all field directions involving $\Phi_i$ and $\Delta$.
	\item[\textbullet]{\bf EWPOs}: The oblique parameters $S$, $T$, and $U$~\cite{Peskin:1991sw,Grimus:2008nb} have been calculated in the 2HDMcT~\cite{Ouazghour:2023eqr}. Using the PDG electroweak inputs~\cite{ParticleDataGroup:2024cfk,ParticleDataGroup:2026aaa} and including the $S$--$T$ correlation, we adopt:	
	\begin{align}
	\widehat S_0=0.008\pm 0.071,~\widehat T_0 = 0.021\pm 0.055,~\rho_{ST} = 0.92 
	\end{align}	
	for which we use the following $\chi^2_{ST}$ test
	
	\begin{equation}
	\small
	\label{eq:STRange}
	\frac{(S-\widehat S_0)^2}{\sigma_S^2}\ +\
	\frac{(T-\widehat T_0)^2}{\sigma_T^2}\ -\
	2\rho_{ST}\frac{(S-\widehat S_0)(T-\widehat T_0)}{\sigma_S \sigma_T}\
	\leq\ R^2\,(1-\rho_{ST}^2)\; ,
	\end{equation}
	with $R^2=2.3$ and $5.99$ corresponding to $68.3 \%$  and
	$95 \% $ Confidence Level (CL), respectively.
	In the numerical analysis, we impose the $95\%$ CL constraint.
	\item \textbf{Colliders}: We also include constraints from collider measurements using the {\texttt{HiggsTools}} package~\cite{Bahl:2022igd}. This ensures that the allowed parameter regions are consistent with the observed properties of the $125$~GeV Higgs boson~(via \texttt{HiggsSignals}~\cite{Bechtle:2013xfa,Bechtle:2014ewa,Bechtle:2020uwn,Bahl:2022igd}) and with the limits from null searches for additional Higgs bosons at the LHC, Tevatron and LEP (via \texttt{HiggsBounds}~\cite{Bechtle:2008jh,Bechtle:2011sb,Bechtle:2013wla,Bechtle:2020pkv,Bahl:2022igd}).
	\item[\textbullet]{\bf Flavor}:  Flavor constraints are also implemented in our analysis. We use the $B$-physics results derived in \cite{Ouazghour:2023eqr}, together with the experimental data at $2\sigma$ CL~\cite{HeavyFlavorAveragingGroupHFLAV:2024ctg} displayed in Tab.~\ref{Tab2}.
\end{itemize}
{\renewcommand{\arraystretch}{1.5}
	{\setlength{\tabcolsep}{0.1cm} 
\begin{table*}[t]
	\centering
	\caption{Experimental constraint on the flavor observable $\mathrm{BR}(\bar{B}\to X_s\gamma)$ at 95\% CL.}
	\label{Tab2}
	\renewcommand{\arraystretch}{1.2}
	\setlength{\tabcolsep}{8pt}
	
	\begin{tabular}{lcc}
		\toprule
		Observable & Experimental result & 95\% CL range \\
		\midrule
		$\mathrm{BR}(\bar{B}\to X_s\gamma)$\cite{Ouazghour:2023eqr}
		& $(3.49\pm0.19)\times10^{-4}$\cite{HeavyFlavorAveragingGroupHFLAV:2024ctg}
		& $[3.11\times10^{-4},\,3.87\times10^{-4}]$ \\
		\bottomrule
	\end{tabular}
\end{table*}
		Constraints specific to the parameter space of the 2HDMcT are as follows:
		\begin{itemize}
			\item \textbf{$\rho$ Parameter}:  
			The additional scalar triplet in the 2HDMcT modifies the $\rho$ parameter as  
			\begin{eqnarray}
			\rho = \frac{v_0^2 + 2v_t^2}{v_0^2 + 4v_t^2} 
			\approx 1 - 2 \frac{v_t^2}{v_0^2} = 1 + \delta\rho.
			\label{eq:rho-thdmt}
			\end{eqnarray}
			In the SM, $\rho = 1$ at tree-level. Any deviation from this value is tightly constrained and the latest global fit to EWPOs~\cite{ParticleDataGroup:2024cfk} yields
			\begin{eqnarray}
			\rho = 1.00031 \pm 0.00019 \, ,
			\end{eqnarray}
			which is about $1.6\sigma$ above the SM tree-level prediction.  
			This small but nonzero deviation imposes an upper bound on the triplet VEV ($v_t$) in the 2HDMcT framework~\cite{Padhan:2019jlc,Ashanujjaman:2021txz}.  
			
			\item \textbf{Lepton Flavor Violation (LFV)}:
			From the Yukawa interaction shown in Eq.\,(\ref{neutrino_mass}), LFV decays such as $\mu\rightarrow e\gamma$ at loop level and $\mu\rightarrow 3e$ at tree level are induced. Their branching ratios (BRs) in the 2HDMcT are given by \cite{Lavoura:2003xp,Kuno:1999jp,Akeroyd:2009nu,Ouazghour:2018mld} 
			\begin{equation}
			{\rm BR}(\mu \rightarrow e \gamma) = 384 \pi^2 \, |A_R|^2
			\end{equation}
			
			\begin{equation}
			A_R = \frac{- q_e |(h^\dagger h)_{e\mu}|}{48 \sqrt{2} \pi^2 G_F}
			\left(
			\frac{C_{23}}{m_{H_1^\pm}^2}
			+ \frac{C_{33}}{m_{H_2^\pm}^2}
			+ \frac{8}{m_{H^{\pm\pm}}^2}
			\right)
			\end{equation}
			
			\begin{equation}
			{\rm BR}(\mu \rightarrow 3e)
			= \frac{|h_{ee}|^2 \, |h_{\mu e}|^2}
			{4 G_F^2 m_{H^{\pm\pm}}^4}
			\end{equation}
			
			with
			\begin{equation}
			\begin{aligned}
			h_{ij}
			&= \frac{m_{ij}}{\sqrt{2} v_t} \\
			&= \frac{1}{\sqrt{2} v_t}
			\left(
			U_{\mathrm{PMNS}}
			\, \mathrm{diag}(m_1, m_2, m_3) \,
			U_{\mathrm{PMNS}}^T
			\right)_{ij}.
			\end{aligned}
			\end{equation}
			
			Here, $\alpha$ refer to the electromagnetic fine-structure constant and $G_F$ denotes the Fermi constant, while $C_{23}$ and $C_{33}$ are elements of the charged-scalar rotation matrix. The explicit expressions for $C_{23}$ and $C_{33}$, as well as for the other rotation matrix elements, can be found in Ref.~\cite{Ouazghour:2018mld}. The upper bounds on the BRs of the above processes are $1.5 \times 10^{-13}$ for $\mu\rightarrow e\gamma$ \cite{MEGII:2025gzr} and $1.0 \times 10^{-12}$ for $\mu\rightarrow 3e$ \cite{SINDRUM:1987nra}. 
			\item[\textbullet]{\bf Neutrino Oscillation Experiments}: 
			The experimental constraints on the neutrino oscillation parameters are given as follows~\cite{Esteban:2024eli} :  
			
			\begin{align}
			& \ \Delta m_{21}^2= \ 7.49_{-0.19}^{+0.19} \times 10^{-5}\ \text{eV}^2, \\
			& \ \sin^2\theta_{12}= \ 0.308_{-0.011}^{+0.012} , \\
			& \sin^2\theta_{13}: 
			\begin{cases}
			\sin^2\theta_{13}= \ 0.02215_{-0.00058}^{+0.00056} \  & \text{(NH)}, \\[4pt]
			\sin^2\theta_{13}= \ 0.02231_{-0.00056}^{+0.00056} \  & \text{(IH)},
			\end{cases} \\[8pt]
			& \sin^2\theta_{23}: 
			\begin{cases}
			\sin^2\theta_{23} = \ 0.470_{-0.013}^{+0.017}  & \text{(NH)}, \\[4pt]
			\sin^2\theta_{23} = \ 0.550_{-0.015}^{+0.012} & \text{(IH)},
			\end{cases} \\[8pt]
			& |\Delta m_{3l}^2|: 
			\begin{cases}
			\ \Delta m_{3l}^2 = \ 2.513_{-0.019}^{+0.021}\times 10^{-3} \ \text{eV}^2 \ \ \text{(NH)}, \\[4pt]
			\ \Delta m_{3l}^2 = \ -2.484_{-0.020}^{+0.020}\times 10^{-3} \ \text{eV}^2 \ \ \text{(IH)},
			\end{cases}
			\end{align}

			Lastly, cosmological observations constrain the sum of neutrino masses to be $\sum_{\nu=1}^3 m_{\nu} \ < 0.12\ \text{eV}$~\cite{ParticleDataGroup:2024cfk}.

		\end{itemize}

		\vspace{-15pt}
		\section{Procedure}
		\label{COMPUTATIONAL_PROCEDURE}
		
To identify the surviving regions of the 2HDMcT parameter space, we perform a parameter scan while enforcing all theoretical and experimental constraints listed above. 
		
The input parameter ranges used are given in Tab.~\ref{tab:input}.
		\begin{table}[t]
			\caption{Input parameter ranges used in the numerical scan. The SM-like Higgs boson is identified with $h_1$. Masses are given in GeV.}
			\label{tab:input}
			\begin{ruledtabular}
				\begin{tabular}{lc}
					Parameter & Scan range \\
					\hline
				$m_{h_1}$ & $125.09$ \\
				$m_{h_2}$ & $[m_{h_1}, \ 1000]$ \\
					$m_{h_3}$ & $[m_{h_2}, \ 1000]$ \\
				$\alpha_1$ & $[-\pi/2\ ,\ \pi/2]$ \\
				$\alpha_{2,3}$ & $[-0.1\ ,\ 0.1]$ \\
				$\tan\beta$ & $[0.5\ ,\ 120]$ \\
				$\mu_1$ & $[-10^2\ ,\ 10^2]$ \\
				$v_t$ & $[0\ ,\ 2]$ \\
				$\lambda_i,\;\bar{\lambda}_i$ & $[-8\pi\ ,\ 8\pi]$ \\
				\end{tabular}
			\end{ruledtabular}
		\end{table}
		
	We identify the lightest CP-even state $h_1$ with the observed Higgs boson and fix its mass to $125.09$~GeV~\cite{ATLAS:2012yve,CMS:2012qbp}. All parameter points are subjected to the constraints described above, and only the surviving points are subsequently passed to~\texttt{FormCalc}~\cite{Hahn:2001rv,Hahn:1998yk,Kublbeck:1990xc}, which we employ to calculate the cross sections for the production channels $\gamma\gamma \to H^{\pm\pm}H_1^{\mp}H_1^{\mp}$ and $\gamma\gamma \to H^{\pm\pm}H_1^{\mp}W^{\mp}$ considered in this work. This ensures that all reported results correspond to parameter points allowed by both theory and experiment.
		
	Overall, the interplay between gauge and scalar interactions determines the hierarchy of the contributing diagrams and governs the behavior of the cross sections within the allowed 2HDMcT parameter space. The Feynman diagrams corresponding to the processes under investigation are shown in Figs.~\ref{Diagram_HppHm1Hm1} and~\ref{Diagram_HppHp1W}.
		\vspace*{-10pt}
		\section{Results}
                     \label{Sec:Results}
		\subsection{Cross sections}
		\begin{figure}[t]  
			\centering
			\includegraphics[scale=0.6]{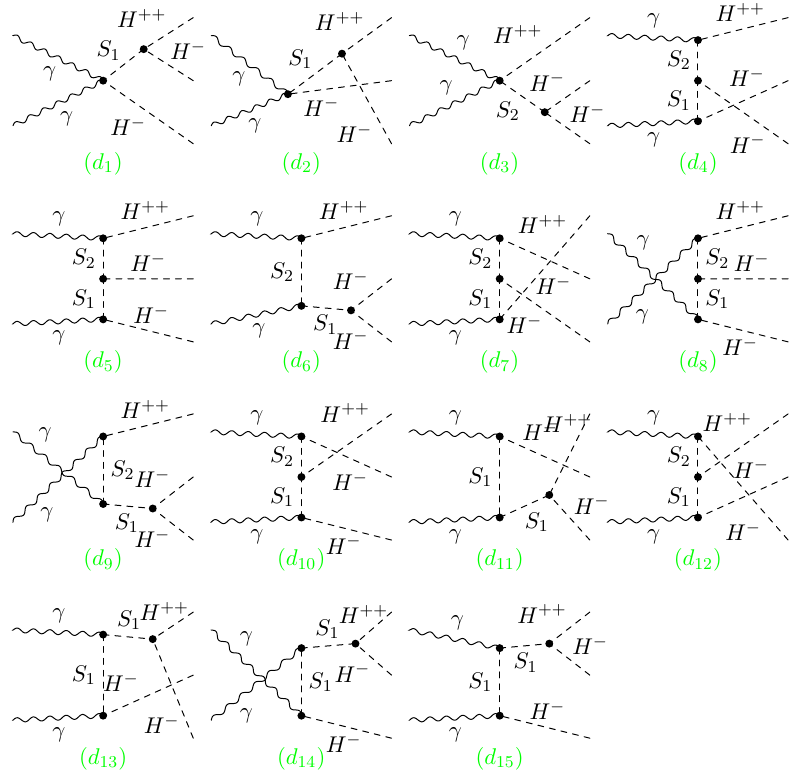}
			\\
			\caption{
				\small
				Tree-level generic Feynman diagrams for $\gamma\gamma\to H^{++}H_1^-H_1^-$ are shown in $(d_{1,...,15})$. For all diagrams $S_1$ refers to $H_1^{\pm}$ and $S_2$ to $H^{\pm\pm}$.}
			\label{Diagram_HppHm1Hm1}
		\end{figure}
		\begin{figure}[t]  
			\centering
			\includegraphics[scale=0.6]{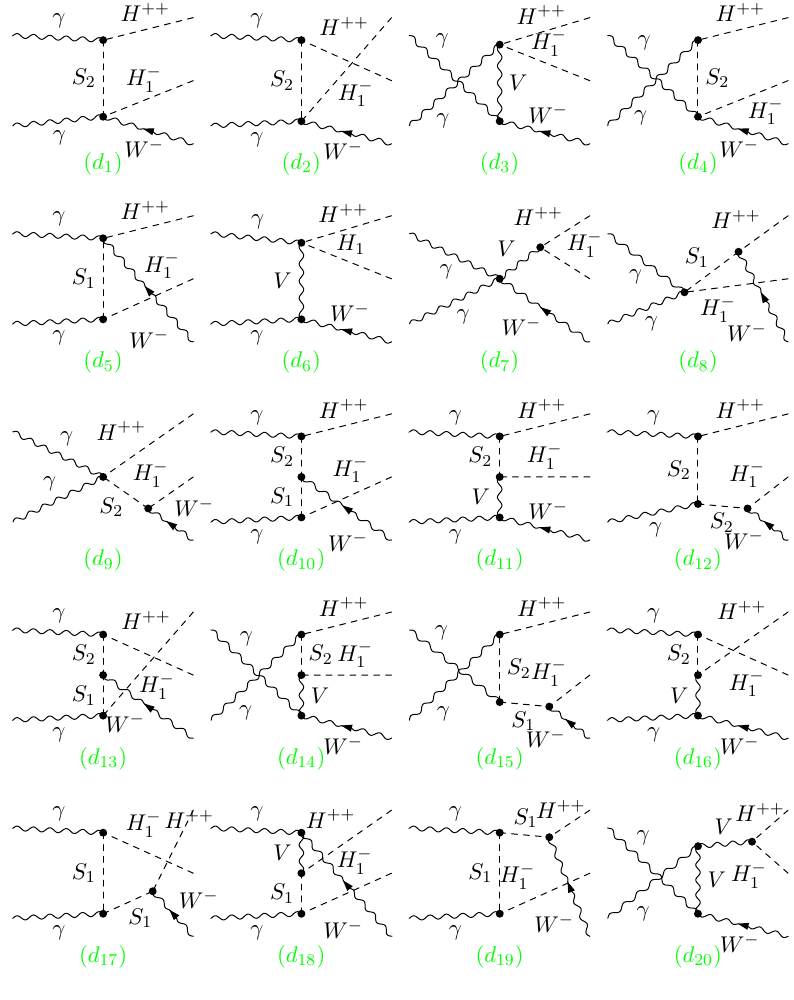}
			\\
			\caption{
				\small
				Tree-level generic Feynman diagrams for $\gamma\gamma\to H^{++}H_1^-W^-$ are shown in $(d_{1,...,20})$. For all diagrams $S_1$ refers to $H_1^{\pm}$ and $S_2$ to $H^{\pm\pm}$, while $V$ refers to $W^{\pm}$.}
			\label{Diagram_HppHp1W}
		\end{figure}
		 As noted above, the mass gap $m_{H^{\pm\pm}}-m_{H_1^{\pm}}$ plays an important role in determining the relative significance of direct three-body production compared with resonant $H^{++}H^{--}$ production followed by $H^{\pm\pm}\to H_1^\pm H_1^\pm$ or $H^{\pm\pm}\to H_1^\pm W^\pm$.
		 
		\begin{figure*}[t]
			\centering
			\includegraphics[scale=0.4]{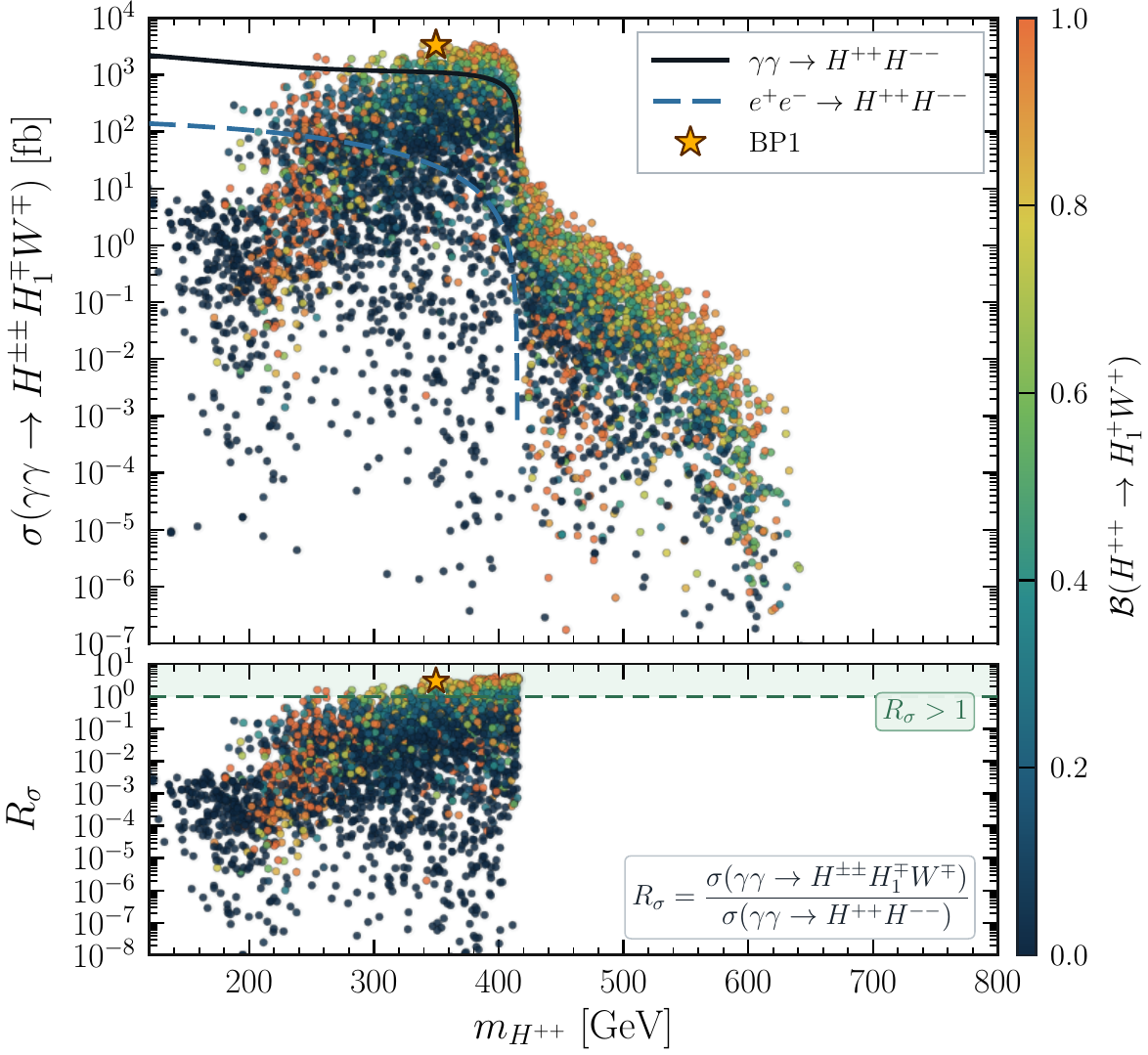}
			\includegraphics[scale=0.4]{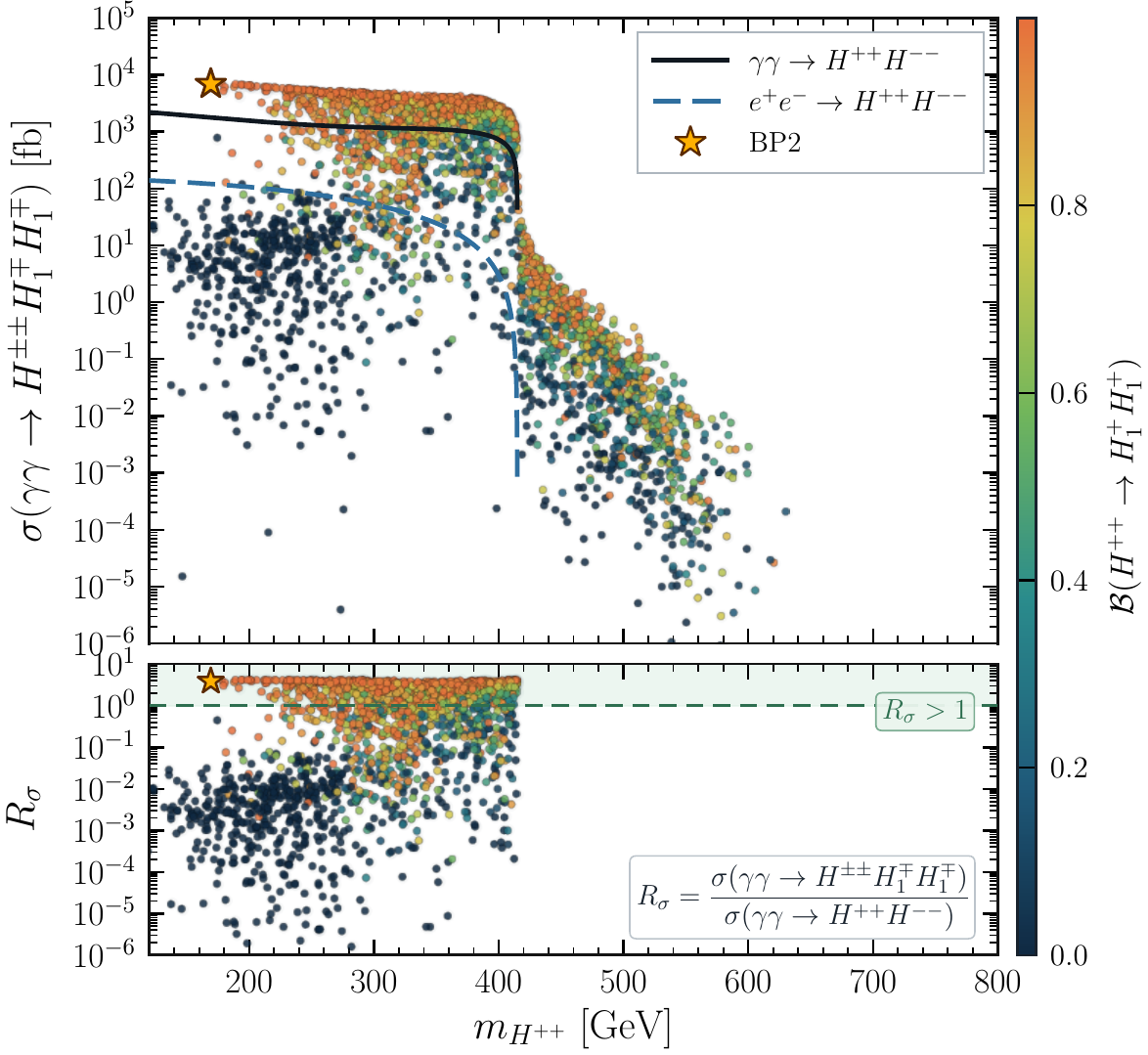}
			\includegraphics[scale=0.4]{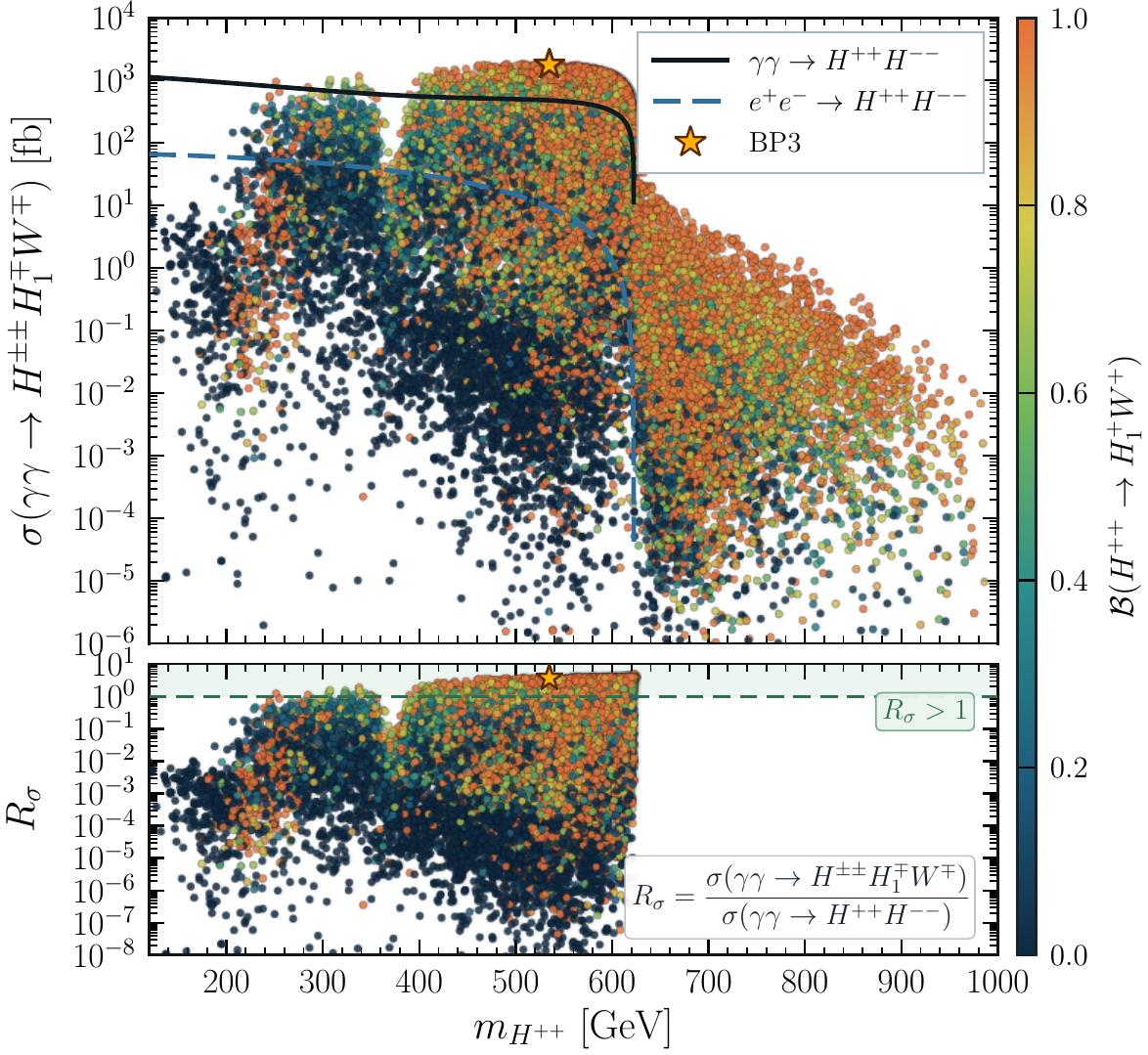}	
			\includegraphics[scale=0.4]{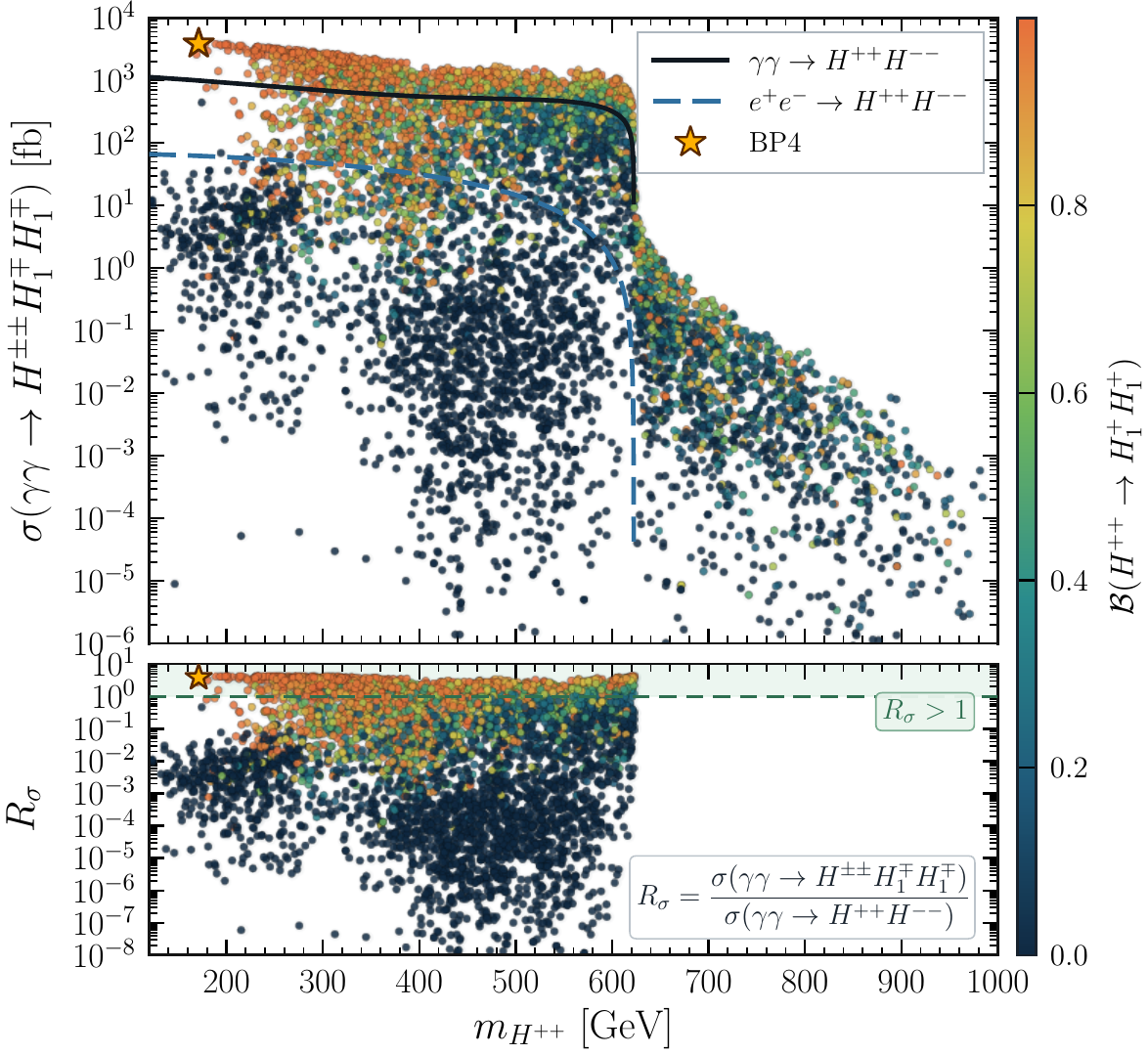}
			\\
			\caption{
				\small
				Production cross sections for $\gamma\gamma \to H^{\pm\pm} H_1^{\mp} H_1^{\mp}$ and $\gamma\gamma \to H^{\pm\pm} H_1^{\mp} W^{\mp}$ as a function of $m_{H^{++}}$ at $\sqrt{s}= 830,\ 1245$ GeV. The green stars indicate BP1-BP4. }
			\label{results_HppHm1Hm1_830}
		\end{figure*}
 We present in Fig.~\ref{results_HppHm1Hm1_830} the cross sections for the processes
$\gamma\gamma \to H^{\pm\pm}H_1^\mp W^\mp$ (left panels) and
$\gamma\gamma \to H^{\pm\pm}H_1^\mp H_1^\mp$ (right panels) at
$\sqrt{s_{\gamma\gamma}}=830$ and $1245~\mathrm{GeV}$, corresponding to $83\%$ of the center-of-mass energies of parent $e^-e^+$ colliders operating at $\sqrt{s}=1000$ and $1500~\mathrm{GeV}$, respectively. For comparison, we also present the cross sections for the pair-production processes $e^-e^+ \to H^{++}H^{--}$ and $\gamma\gamma \to H^{++}H^{--}$, represented by the blue and black curves, respectively.
These production channels are primarily governed by the trilinear scalar vertex $H^{++}H_1^-H_1^-$ and the gauge-Higgs vertex $H^{++}H_1^-W^-$, which are proportional to the couplings $\lambda_{H^{++}H_1^-H_1^-}$ and $C_{23}$, respectively.  
Consequently, the majority of the contributing diagrams have a strong dependence on these interactions.  

Figure~\ref{results_HppHm1Hm1_830} shows that the pair-production cross section $\gamma\gamma \to H^{++}H^{--}$ is more than an order of magnitude larger than the corresponding $e^-e^+ \to H^{++}H^{--}$ rate at the same center-of-mass energy\footnote{A similar enhancement is also observed for the $2\to3$ processes.}. For the $H^{\pm\pm}H_1^\mp H_1^{\mp}$ final state, the dominant contribution often arises from $H^{++}H^{--}$ production followed by $H^{\pm\pm}\to H_1^\pm H_1^\pm$. Nevertheless, the remaining diagrams can give sizable contributions both when this decay is kinematically forbidden, $m_{H^{--}}<2m_{H_1^-}$, and when it is open. The full $2\to3$ rate can therefore exceed the factorized estimate $\sigma(\gamma\gamma \to H^{++}H^{--})\times {\rm BR}(H^{\pm\pm}\to H_1^\pm H_1^\pm)$ by more than a factor of 2, as indicated by values of $R_{\sigma}>2$ in the region $m_{H^{--}}>2m_{H_1^-}$ shown in Fig.~\ref{results_HppHm1Hm1_830}. A similar trend is observed for the $H^{\pm\pm}H_1^\mp W^{\mp}$ final state.
\begin{table}[!h]
	\centering
	\caption{BPs used for the signal-to-background analysis.}
	\label{BPs}
	\renewcommand{\arraystretch}{1.0}
	\setlength{\tabcolsep}{-0.8pt}
	\small
	\begin{tabular}{lcccc}
		\toprule
		Parameter & BP1 & BP2 & BP3 & BP4 \\
		\midrule
		$m_{h_1}$            & 125.09 & 125.09 & 125.09 & 125.09 \\
		$m_{h_2}$            & 142.15 & 130.26 & 160.41 & 145.30 \\
		$\lambda_1$          & 0.9611 & 0.4930 & 0.9773 & 0.5678 \\
		$\lambda_3$          & 0.1073 & $-8.95\times10^{-2}$ & $-0.1755$ & $-4.03\times10^{-2}$ \\
		$\lambda_4$          & $2.49\times10^{-2}$ & 0.2280 & 0.7280 & 0.2655 \\
		$\lambda_6$          & 1.5638 & 0.4748 & 0.4919 & 0.7227 \\
		$\lambda_7$          & $1.45\times10^{-2}$ & $4.03\times10^{-2}$ & 0.2841 & $7.99\times10^{-2}$ \\
		$\lambda_8$          & $-0.7733$ & 0.1857 & $-0.4205$ & $-0.5346$ \\
		$\lambda_9$          & 0.5370 & $8.74\times10^{-2}$ & 0.5829 & $7.68\times10^{-2}$ \\
		$\bar{\lambda}_8$    & 1.1199 & 1.6795 & 2.7661 & 2.0904 \\
		$\bar{\lambda}_9$    & 2.4684 & $-0.5696$ & $-0.1189$ & 0.3526 \\
		$\tan\beta$          & 82.45 & 79.19 & 119.48 & 92.58 \\
		$\alpha_1$           & 1.5597 & 1.5586 & 1.5629 & $-1.5615$ \\
		$\alpha_2$           & $-4.62\times10^{-4}$ & $-5.01\times10^{-4}$ & $-3.11\times10^{-4}$ & $-2.96\times10^{-4}$ \\
		$\alpha_3$           & $-4.02\times10^{-2}$ & 0.1025 & $9.30\times10^{-2}$ & $-3.30\times10^{-2}$ \\
		$v_t$                & 0.3843 & 0.2684 & 0.4775 & $5.88\times10^{-2}$ \\
		\bottomrule
	\end{tabular}
\end{table}
\subsection{Signal-to-background analysis}
To assess the observability of the doubly charged Higgs boson $H^{\pm\pm}$, particularly in regions where the full $2\to3$ processes are important, we develop search strategies aimed at maximizing the separation between signal and SM backgrounds. The analysis relies on a simulation chain that accounts for matrix-element generation, resonance decays, parton showering and hadronization, heavy flavor decays, jet reconstruction, and detector effects. We focus on the clean multilepton signature $4\ell+\slashed{E}_T$ as a representative case study, which we will show to provide strong signal sensitivity when the dominant backgrounds are efficiently suppressed. For each center-of-mass energy considered, we select a representative BP from the scan, reported in Tab.~\ref{BPs}. 

We consider center-of-mass energies of $\sqrt{s_{\gamma\gamma}}=830$ and $1245~\mathrm{GeV}$, corresponding to approximately $83\%$ of the center-of-mass energies of parent $e^-e^+$ colliders operating at $\sqrt{s}=1000$ and $1500~\mathrm{GeV}$, respectively. As discussed above, we focus on the high-energy peak of the photon spectrum and conservatively assume that the effective $\gamma\gamma$ luminosity amounts to $10\%$ of the parent $e^-e^+$ collider luminosity. Accordingly, we consider integrated luminosities of $50$, $100$, and $150~\mathrm{fb}^{-1}$, corresponding to $10\%$ of parent $e^-e^+$ integrated luminosities of $500$, $1000$, and $1500~\mathrm{fb}^{-1}$, respectively. Although these effective luminosities are significantly reduced, this limitation is largely compensated by the substantially enhanced production cross sections of the signal processes, which exceed their corresponding $e^-e^+$ production cross sections by more than one order of magnitude.

For $4\ell+\slashed{E}_T$ signature, the dominant SM backgrounds primarily stem from multiboson production:
$W^+W^-Z$, $W^+W^-\gamma$, $W^+W^-ZZ$ and $W^+W^-W^+W^-$. Here, we assume that the $W^\pm$ bosons decay leptonically. The $Z$ bosons decay into $\ell^+\ell^-$ or $\nu_\ell\bar\nu_\ell$, while $\gamma^*\to \ell^+\ell^-$ is imposed when relevant. The relevant backgrounds considered in our analysis are
\small
\begin{align}
	\gamma\gamma &\rightarrow W^+W^-Z,
	& W^-&\rightarrow \ell^- \bar{\nu}_{\ell},~
	W^+\rightarrow \ell^+\nu_{\ell},~
	Z\rightarrow \ell^+\ell^- ,
	\\[0mm]
	\gamma\gamma &\rightarrow W^+W^-\gamma,
	& W^-&\rightarrow \ell^- \bar{\nu}_{\ell},~
	W^+\rightarrow \ell^+\nu_{\ell},~
	\gamma^*\rightarrow \ell^+\ell^- ,
	\\[0mm]
	\gamma\gamma &\rightarrow W^+W^-W^+W^-,
	& W^\pm&\rightarrow \ell^\pm\nu_{\ell},
	\\[0mm]
	\gamma\gamma &\rightarrow W^+W^-ZZ,
	& W^\pm&\rightarrow \ell^\pm\nu_{\ell},~
	Z\rightarrow \ell^+\ell^-,~
	Z\rightarrow \nu_{\ell}\bar{\nu}_{\ell}.
\end{align}
\normalsize	

Events are required to satisfy the following baseline acceptance and isolation criteria:
\begin{equation}
	|\eta^{j,\ell}| < 2.5,\qquad
	\Delta R^{\ell\ell} \geq 0.4,\qquad
	P_{T}^{\ell} \geq 10 \, GeV,
	\label{Bcuts}
\end{equation}
which ensure that the reconstructed leptons and jets lie within the detector acceptance and are sufficiently well separated.
		\begin{figure*}[t]
			\centering	
			\includegraphics[scale=0.29]{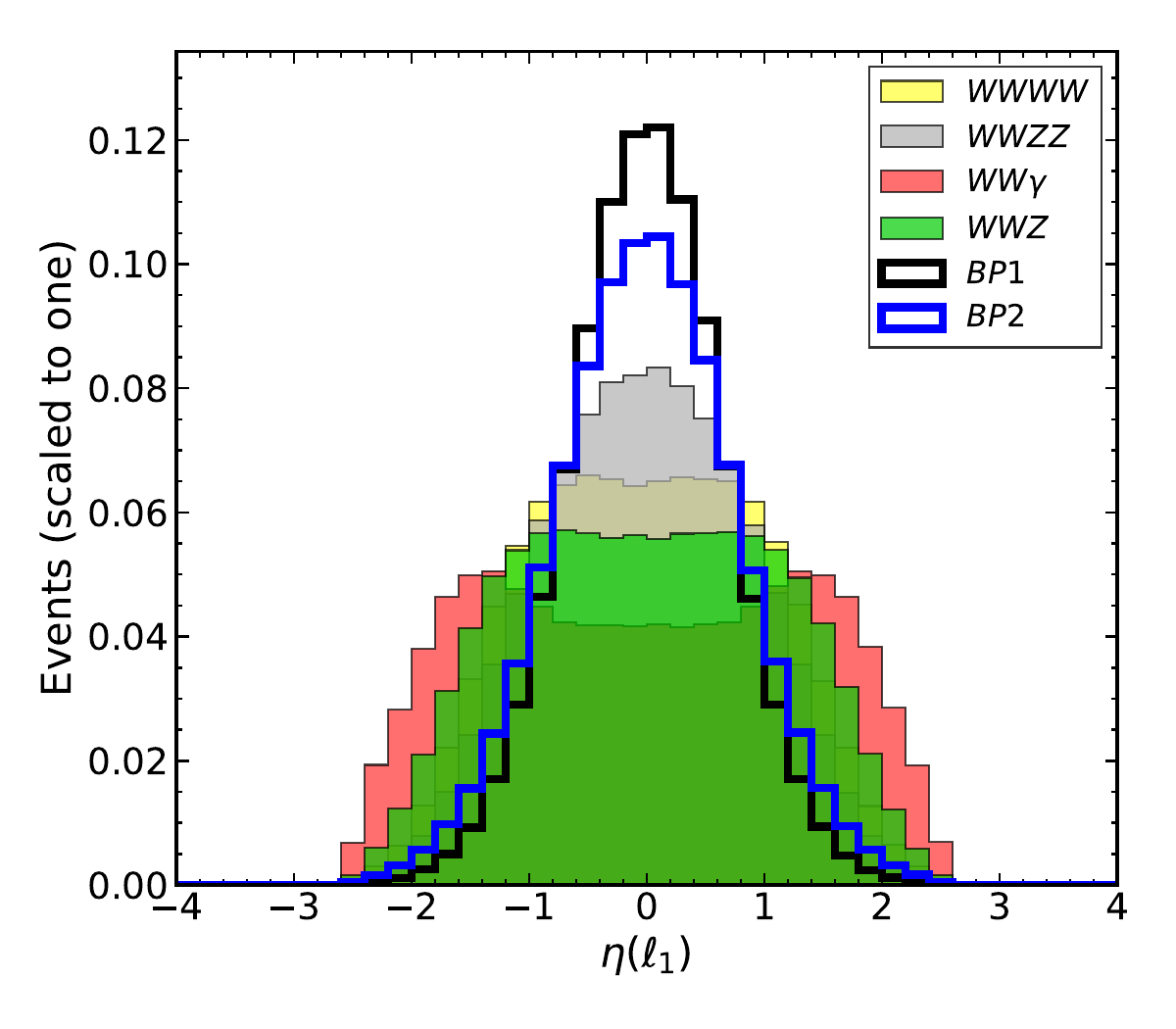}
			\includegraphics[scale=0.29]{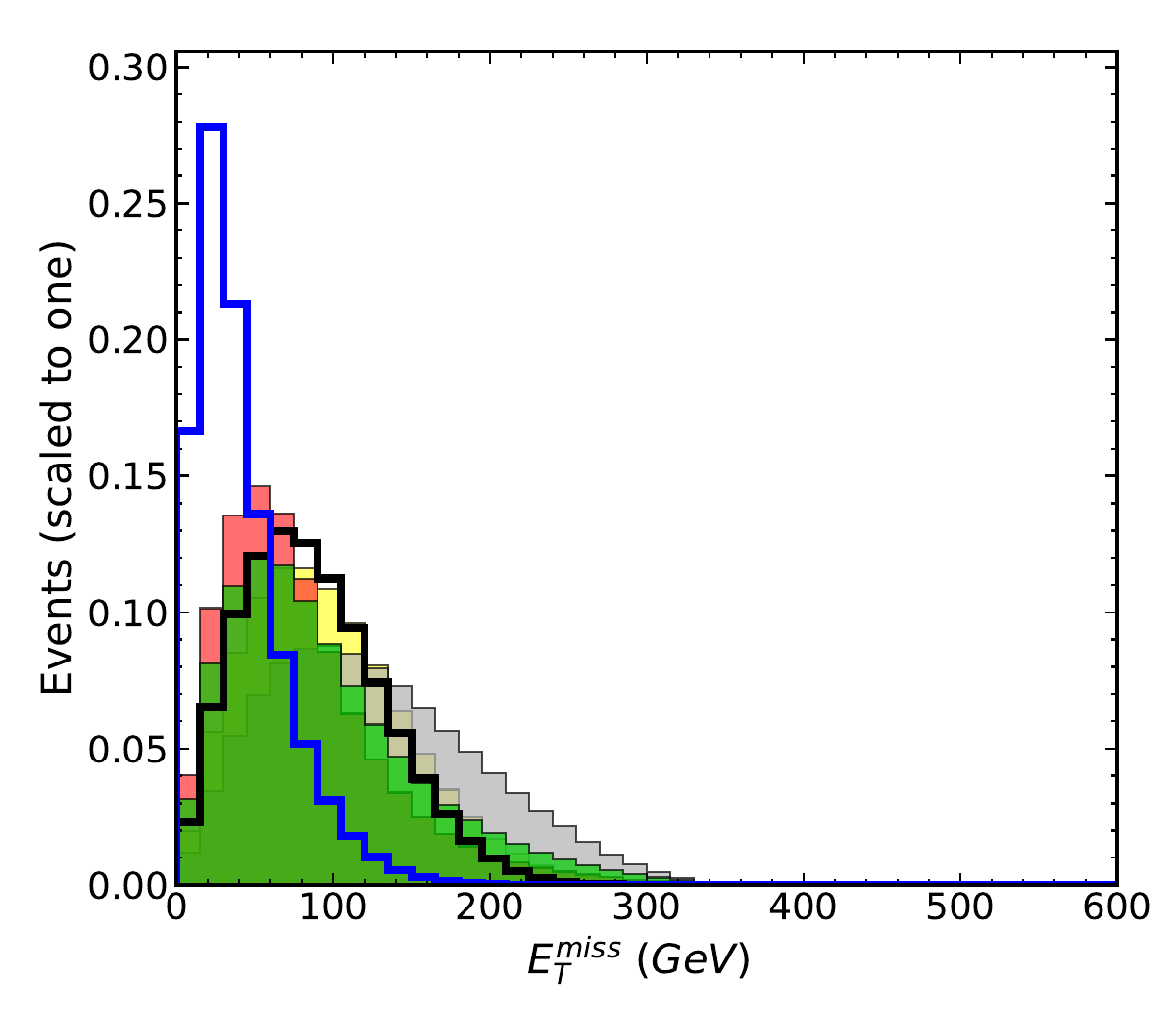}
			\includegraphics[scale=0.29]{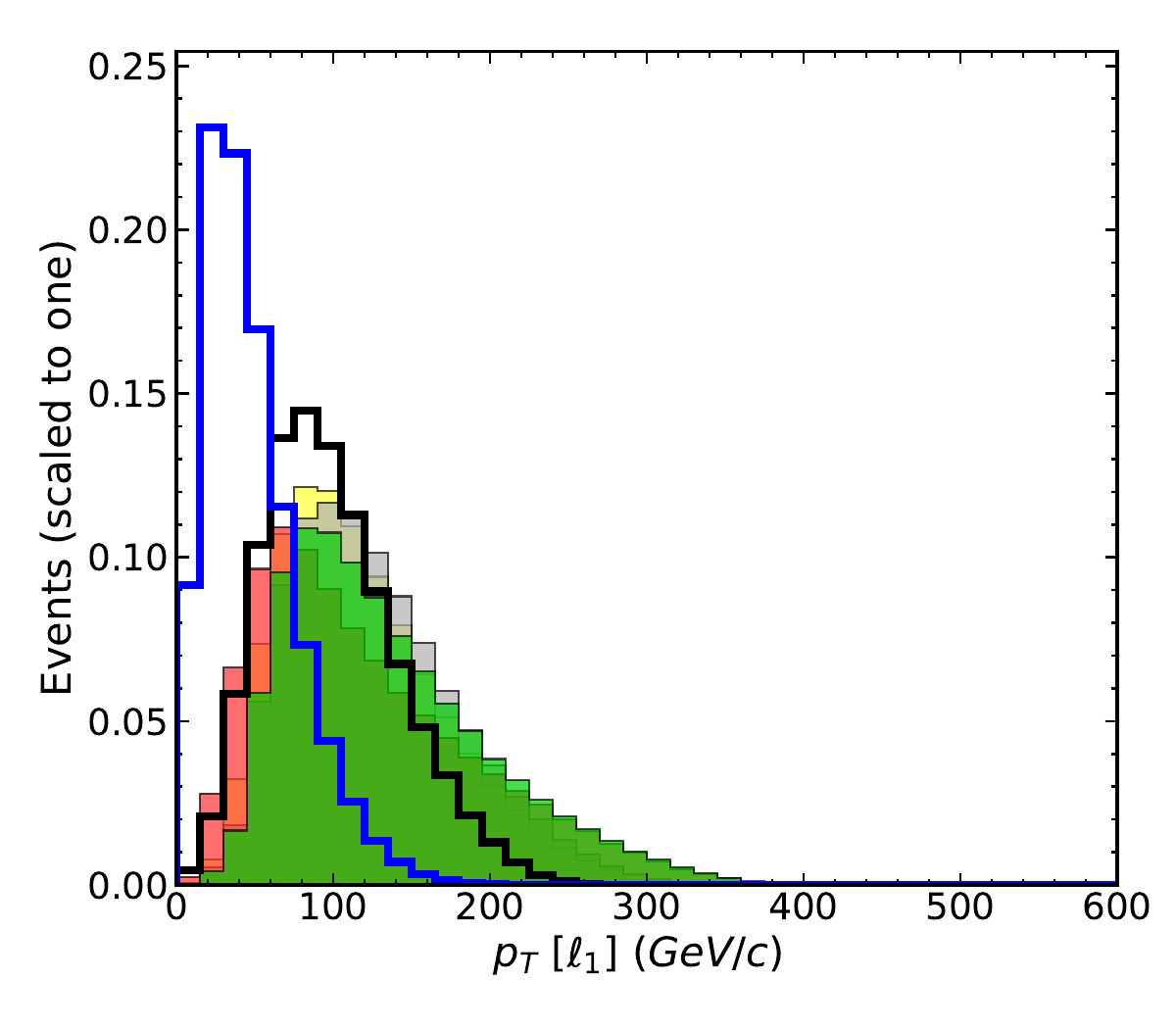}	
			\includegraphics[scale=0.29]{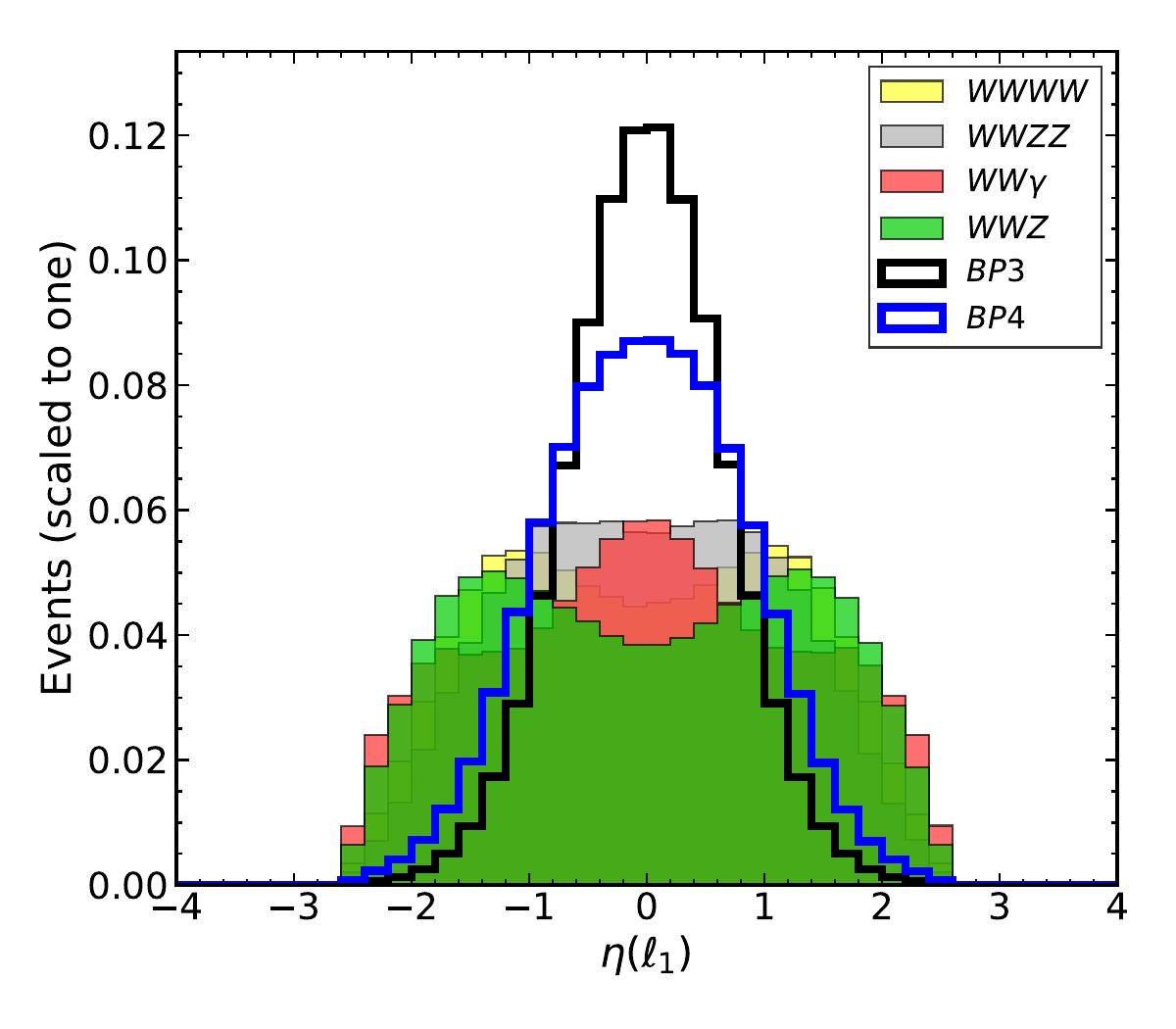}
			\includegraphics[scale=0.29]{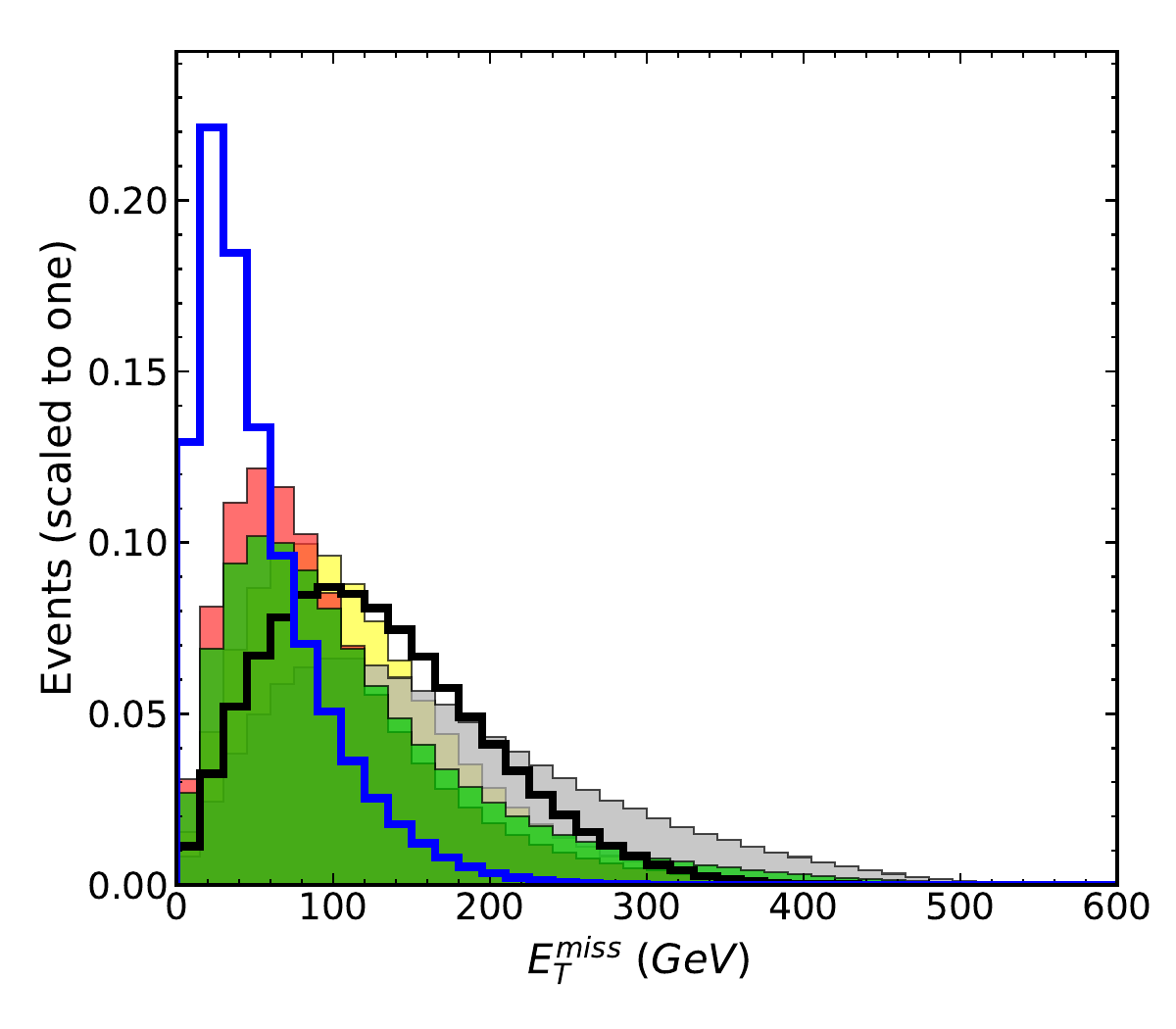}
			\includegraphics[scale=0.29]{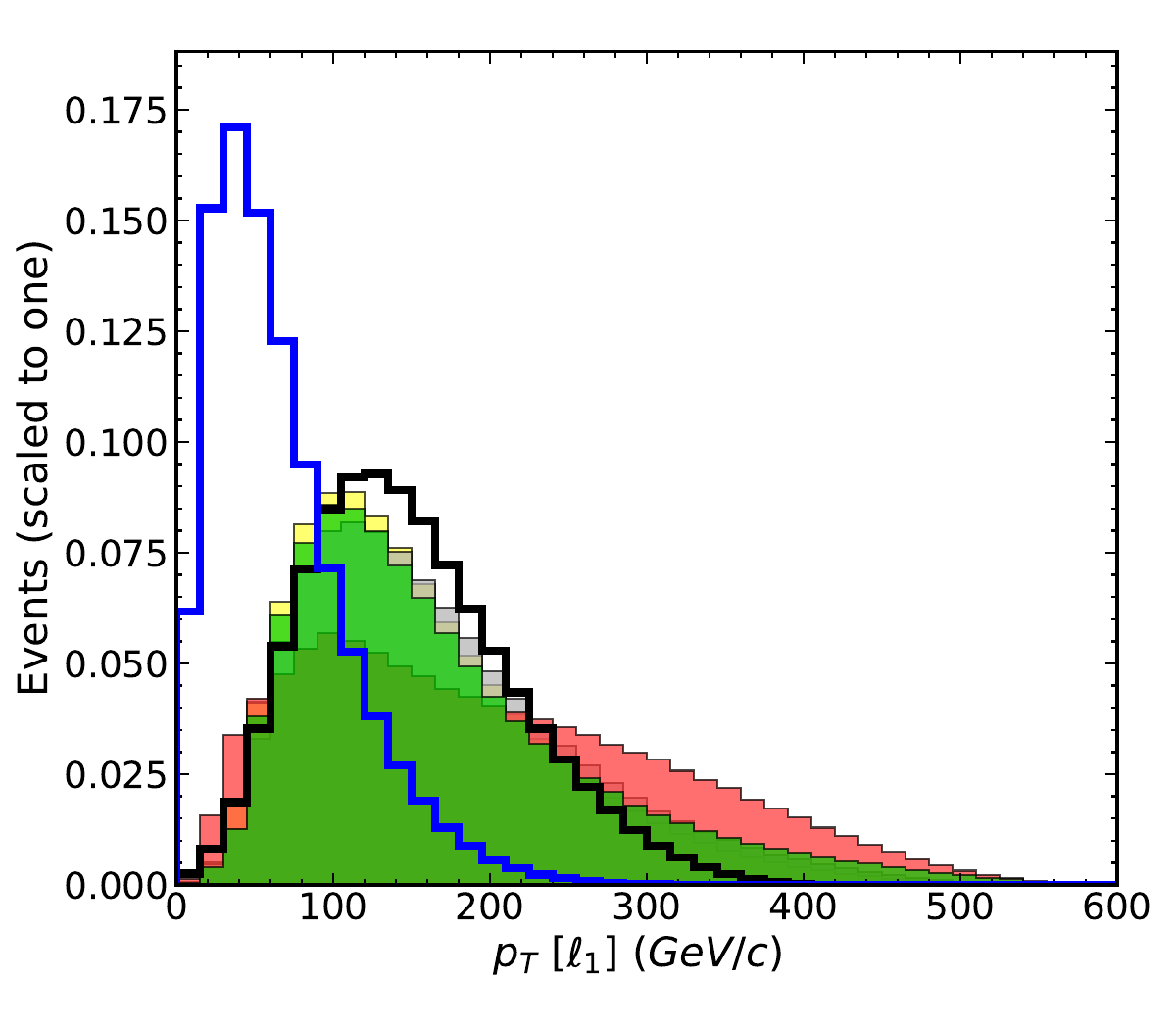}
			\caption{Kinematic distributions for the $4\ell + \slashed{E}_T$ final state at $\sqrt{s}=830$ GeV~(upper panels) and $\sqrt{s}=1245$ GeV~(lower panels). The pseudorapidity $\eta(\ell_1)$ of the leading lepton (left), the missing transverse energy $\slashed{E}_T$ (middle) and the transverse momentum $p_T(\ell_1)$ (right).}
			\label{dist}
		\end{figure*}
      We use the median discovery significance of Ref.~\cite{Cowan:2010js} to quantify the signal sensitivity $\mathcal{Z}$. For signal and background yields $s$ and $b$, respectively, and a fractional systematic uncertainty $\delta$ on the background, we compute
		\small
			\begin{align}
			\mathcal{Z} =
			\sqrt{2\Bigg[\left(s+b\right)\ln\!\left(\frac{\left(s+b\right)\left(1+\delta^2b\right)}{b+\delta^2b\left(s+b\right)}\right)-\frac{1}{\delta^2}\ln\!\left(1+\delta^2\frac{s}{1+\delta^2b}\right)\Bigg]}
			\end{align}
		\normalsize

Signal events are generated at parton level using~\texttt{MadGraph5\_aMC\_v3.4.2}~\cite{Alwall:2014hca,Hagiwara:2012vz}. The generated events are subsequently passed to \texttt{Pythia-8.20}~\cite{Sjostrand:2007gs} for parton showering,  hadronization, and heavy flavor decays. Jets are clustered using \texttt{FastJet}~\cite{Cacciari:2011ma}, while detector effects are simulated with \texttt{Delphes-3.4.5}~\cite{deFavereau:2013fsa}, implementing the anti-$k_t$ algorithm~\cite{Cacciari:2008gp} with cone size $R=0.5$. The resulting signal and SM background samples are finally processed and analyzed with \texttt{MadAnalysis5}~\cite{Conte:2012fm,Conte:2013mea}.

The $4\ell+\slashed{E}_T$ final state can arise through two distinct signal topologies. The first originates from the associated production of $H^{\pm\pm}H_1^\mp W^\mp$, where the doubly charged Higgs boson decays via $H^{\pm\pm}\to H_1^\pm W^\pm$, followed by the leptonic decay of the singly charged Higgs boson, $H_1^\pm\to\tau_\ell\nu_\tau$, and the leptonic decay of the $W$ bosons :
\begin{eqnarray}
	\gamma\gamma \to H^{\pm\pm} H_1^\mp W^{\mp}
	&\to&
	H_1^{\pm}W^{\pm}\,H_1^\mp W^{\mp}
	\nonumber\\
	&\to&
	\tau_\ell^+\,\ell^+\,\tau_\ell^-\,\ell^-+\slashed{E}_T
	\nonumber\\
	&\to&
	4\ell+\slashed{E}_T\, .
	\nonumber
\end{eqnarray}

The second topology is provided by the production process $\gamma\gamma\to H^{\pm\pm}H_1^\mp H_1^\mp$. In this case, the doubly charged Higgs boson decays into a pair of singly charged Higgs bosons, each of which subsequently decays through the leptonic $\tau$ channel :
\begin{eqnarray}
	\gamma\gamma \to H^{\pm\pm} H_1^\mp H_1^\mp
	&\to&
	H_1^{\pm}H_1^{\pm}H_1^\mp H_1^\mp
	\nonumber\\
	&\to&
	\tau_\ell^+\tau_\ell^+\tau_\ell^-\tau_\ell^-+\slashed{E}_T
	\nonumber\\
	&\to&
	4\ell+\slashed{E}_T\, .
	\nonumber
\end{eqnarray}

Throughout this analysis, $\ell=e,\mu$ denotes a light charged lepton, while $\tau_\ell$ represents the leptonic decay of the $\tau$ lepton. The missing transverse energy, $\slashed{E}_T$, is generated by the neutrinos emitted in the decays of the $\tau$ leptons and the $W^\pm$ bosons.

\subsubsection{Cut strategy}
Fig.~\ref{dist} displays the kinematic distributions of the $4\ell + \slashed{E}_T$ final state at a center-of-mass energy of $\sqrt{s}=830$ GeV (upper panels) and $\sqrt{s}=1245$ GeV (lower panels). Shown are the pseudorapidity of the leading lepton, $\eta(\ell_1)$, in the left panel, the missing transverse energy, $\slashed{E}_T$, in the middle panel, and the transverse momentum of the leading lepton, $p_T(\ell_1)$, in the right panel.

Guided by the kinematic features discussed above, we apply a sequential cut strategy for each BP, as summarized in Tab.~\ref{selection_cuts}. For BP1, we illustrate in detail the impact of each selection requirement on both the signal and background processes. Requiring the leading lepton to lie in the central region, $-1.6<\eta(\ell_1)<1.6$, suppresses the $W^+W^-(Z/\gamma)$ background by about $15\%$, while preserving approximately $98.1\%$ of the signal. The additional requirement $p_T(\ell_1)<200$~GeV further reduces the background contribution by around $20\%$ with only a small loss in signal efficiency. Finally, imposing $\slashed{E}_T<200$~GeV provides a further reduction of about $15\%$ in the $W^+W^-ZZ$ background. Overall, these selection criteria lead to a reduction of approximately $35\%$ in the total background cross section, while affecting the signal by only about $4.9\%$. The cumulative effect of the different cuts on the final cross sections is summarized in Tab.~\ref{cutflow_BP1_BP4}.

For the remaining BPs, a similar selection strategy is adopted, and the corresponding cut-flow results are summarized in Tab.~\ref{cutflow_BP1_BP4}. The overall impact of these requirements remains favorable, reducing the background cross section by about $32\%$, $69\%$, and $31\%$ for BP2, BP3, and BP4, respectively, while decreasing the signal cross sections by only about $1.19\%$, $17.9\%$, and $2.3\%$.

\subsubsection{Significance}
In Tab.~\ref{Significance_AllBP}, we show the expected discovery significance $\mathcal{Z}$ for all BPs. The analysis includes systematic uncertainties of $\delta=5\%$ and $10\%$ at center-of-mass energies of $\sqrt{s}=830$ and $1245$~GeV. Since the effective $\gamma\gamma$ luminosity is assumed to represent $10\%$ of the luminosity of the parent $e^-e^+$ collider, we consider integrated $\gamma\gamma$ luminosities of $50$, $100$, and $150~\mathrm{fb}^{-1}$, corresponding to parent collider luminosities of $500$, $1000$, and $1500~\mathrm{fb}^{-1}$, respectively. 

The results indicate that both processes, $\gamma\gamma \to H^{++} H_1^- W^- \to 4\ell + \slashed{E}_T$ and $\gamma\gamma \to H^{++} H_1^- H_1^- \to 4\ell + \slashed{E}_T$, can reach discovery-level sensitivity, with the latter channel providing larger significances as a consequence of its enhanced production cross section. As expected, increasing the assumed systematic uncertainty leads to a moderate decrease in $\mathcal{Z}$, while moving to higher center-of-mass energies results in a more pronounced loss of sensitivity due to the reduction of the signal production rates. 

The obtained sensitivities show that $\gamma\gamma$ collisions provide a promising complementary avenue for exploring doubly charged Higgs bosons. Despite the smaller effective luminosity considered here, the resulting significances remain comparable or even larger than those obtained in $e^+e^-$ collisions~(see Ref.~\cite{BrahimAit-Ouazghour:2026stq}). This can be traced back to the much larger production rates in the $\gamma\gamma$ mode, where the direct QED couplings of the doubly charged scalar lead to cross sections enhanced by more than one order of magnitude.

		\begin{table}[!h]
			\centering
			\renewcommand{\arraystretch}{1.3}
			\setlength{\tabcolsep}{35pt}
			\begin{adjustbox}{max width=\textwidth}
				\begin{tabular}{lc}
					\hline\hline
					\textbf{Cuts} & \textbf{Definition} \\
					\hline\hline
					
					\multicolumn{2}{c}{\textbf{BP1} ($\sqrt{s}=830$ GeV)}\\
					\hline
					\textbf{Cut-1} & $-1.6 < \eta[l_{1}] < 1.6$ \\
					\hline
					\textbf{Cut-2} & $P_T[l_{1}] < 200$ GeV \\
					\hline
					\textbf{Cut-3} & $\slashed{E}_T < 200$ GeV \\
					\hline
					
					\multicolumn{2}{c}{\textbf{BP2} ($\sqrt{s}=830$ GeV)}\\
					\hline
					\textbf{Cut-1} & $P_T[l_{1}] < 180$ GeV \\
					\hline
					\textbf{Cut-2} & $\slashed{E}_T < 150$ GeV \\
					\hline
					
					\multicolumn{2}{c}{\textbf{BP3} ($\sqrt{s}=1245$ GeV)}\\
					\hline
					\textbf{Cut-1} & $-0.9 < \eta[l_{1}] < 0.9$ \\
					\hline
					\textbf{Cut-2} & $P_T[l_{1}] < 350$ GeV \\
					\hline
					\textbf{Cut-3} & $\slashed{E}_T < 350$ GeV \\
					\hline
					
					\multicolumn{2}{c}{\textbf{BP4} ($\sqrt{s}=1245$ GeV)}\\
					\hline
					\textbf{Cut-1} & $P_T[l_{1}] < 220$ GeV \\
					\hline
					\textbf{Cut-2} & $\slashed{E}_T < 220$ GeV \\
					\hline\hline
					
				\end{tabular}
			\end{adjustbox}
			\caption{Selection criteria used in the signal-to-background analysis for the BPs.}
			\label{selection_cuts}
		\end{table}
		
		\begin{table*}[t]
			\centering
			\setlength{\tabcolsep}{15pt}
			\renewcommand{\arraystretch}{1.2}
			
			\begin{tabular}{lccccc}
				\hline\hline
				
				\multicolumn{6}{c}{\textbf{BP1}}\\
				\hline
				Cuts & Signal & $W^+W^-Z$ & $W^+W^-\gamma$ & $W^+W^-ZZ$ & $W^+W^-W^+W^-$\\
				\hline
				
				Basic cut
				& $4.49$
				& $2.485$
				& $1.946\times10^{-1}$
				& $6.06\times10^{-3}$
				& $5.00\times10^{-2}$ \\
				
				Cut-1
				& $4.404$
				& $2.12$
				& $1.40\times10^{-1}$
				& $5.71\times10^{-3}$
				& $4.54\times10^{-2}$ \\
				
				Cut-2
				& $4.319$
				& $1.71$
				& $1.10\times10^{-1}$
				& $5.08\times10^{-3}$
				& $4.10\times10^{-2}$ \\
				
				Cut-3
				& $4.270$
				& $1.615$
				& $1.07\times10^{-1}$
				& $4.41\times10^{-3}$
				& $3.99\times10^{-2}$ \\
				\textbf{Total efficiencies (\%)}
				& \textbf{95.1}
				& \textbf{65.0}
				& \textbf{55.1}
				& \textbf{72.8}
				& \textbf{79.8} \\
				
				\hline
				\multicolumn{6}{c}{\textbf{BP2}}\\
				\hline
				
				Basic cut
				& $20.48$
				& $2.485$
				& $1.946\times10^{-1}$
				& $6.06\times10^{-3}$
				& $5.00\times10^{-2}$ \\
				
				Cut-1
				& $20.34$
				& $1.925$
				& $1.55\times10^{-1}$
				& $5.06\times10^{-3}$
				& $4.30\times10^{-2}$ \\
				
				Cut-2
				& $20.24$
				& $1.693$
				& $1.45\times10^{-1}$
				& $3.62\times10^{-3}$
				& $3.81\times10^{-2}$ \\
				
				\textbf{Total efficiencies (\%)}
				& \textbf{98.81}
				& \textbf{68.10}
				& \textbf{74.7}
				& \textbf{59.70}
				& \textbf{76.20}  \\
				
				\hline
				\multicolumn{6}{c}{\textbf{BP3}}\\
				\hline
				
				Basic cut
				& $3.06$
				& $3.349$
				& $3.053\times10^{-1}$
				& $1.9154\times10^{-2}$
				& $1.384\times10^{-1}$\\
				
				Cut-1
				& $2.541$
				& $1.259$
				& $1.41\times10^{-1}$
				& $9.93\times10^{-3}$
				& $5.98\times10^{-2}$ \\
				
				Cut-2
				& $2.530$
				& $1.05$
				& $1.04\times10^{-1}$
				& $9.08\times10^{-3}$
				& $5.54\times10^{-2}$ \\
				
				Cut-3
				& $2.521$
				& $1.01$
				& $1.02\times10^{-1}$
				& $8.31\times10^{-3}$
				& $5.50\times10^{-2}$ \\
				
				\textbf{Total efficiencies (\%)}
				& \textbf{82.10}
				& \textbf{30.10}
				& \textbf{33.60}
				& \textbf{43.40}
				& \textbf{39.80} \\
				
				\hline
				\multicolumn{6}{c}{\textbf{BP4}}\\
				\hline
				
				Basic cut
				& $6.84$
				& $3.349$
                & $3.053\times10^{-1}$
                & $1.9154\times10^{-2}$
                & $1.384\times10^{-1}$\\
				
				Cut-1
				& $6.695$
				& $2.526$
				& $1.85\times10^{-1}$
				& $1.45\times10^{-2}$
				& $1.10\times10^{-1}$ \\
				
				Cut-2
				& $6.683$
				& $2.355$
				& $1.80\times10^{-1}$
				& $1.18\times10^{-2}$
				& $1.06\times10^{-1}$ \\
				
				\textbf{Total efficiencies (\%)}
				& \textbf{97.70}
				& \textbf{70.00}
				& \textbf{58.90}
				& \textbf{61.30}
				& \textbf{76.80} \\
				
				\hline\hline
			\end{tabular}
			
			\caption{Cut-flow of the signal and SM background cross sections (fb) after successive cuts for  BP1--BP4 at $\sqrt{s}=830, \ 1245$~GeV.}
			\label{cutflow_BP1_BP4}
		\end{table*}
\vspace*{-6pt}

	\begin{table}[t]
		\centering
		\renewcommand{\arraystretch}{1.2}
		\setlength{\tabcolsep}{8pt}
		\begin{tabular}{c c c c}
			\hline\hline
			\textbf{BP} & $\mathcal{L}$ [fb$^{-1}$] & $\delta=5\%$ & $\delta=10\%$ \\
			\hline
			
			\multirow{3}{*}{BP1 ($\sqrt{s}=830$ GeV)}
			& 50  & 13.02 & 9.30  \\
			& 100 & 16.01 & 10.24 \\
			& 150 & 17.60 & 10.63 \\
			\hline
			
			\multirow{3}{*}{BP2 ($\sqrt{s}=830$ GeV)}
			& 50  & 40.41 & 27.34 \\
			& 100 & 48.25 & 29.61 \\
			& 150 & 52.23 & 30.52 \\
			\hline
			
			\multirow{3}{*}{BP3 ($\sqrt{s}=1245$ GeV)}
			& 50  & 7.92 & 5.47 \\
			& 100 & 9.58 & 5.94 \\
			& 150 & 10.42 & 6.12 \\
			\hline
			
			\multirow{3}{*}{BP4 ($\sqrt{s}=1245$ GeV)}
			& 50  & 15.58 & 10.51 \\
			& 100 & 18.61 & 11.33 \\
			& 150 & 20.11 & 11.65 \\
			\hline\hline
			
		\end{tabular}
		
		\caption{Statistical significances $\mathcal{Z}$ for all BPs at center-of-mass energies $\sqrt{s}=830$ and $1245$ GeV for integrated luminosities of $\mathcal{L}=50$, $100$, and $150~\mathrm{fb}^{-1}$, including systematic uncertainties of $\delta=5\%$ and $10\%$.}
		\label{Significance_AllBP}
	\end{table}
		\section{CONCLUSIONS}
		\vspace{6pt}
High-energy $\gamma\gamma$ collisions, realizable as an operational mode of future linear colliders such as the ILC and CLIC, provide a promising environment for probing extended Higgs sectors.

We investigated the sensitivity of such collisions to doubly charged Higgs bosons within the 2HDM with type-II seesaw (2HDMcT). Focusing on the three-body production channels $\gamma\gamma \to H^{\pm\pm}H_1^{\mp}H_1^{\mp}$ and $\gamma\gamma \to H^{\pm\pm}H_1^{\mp}W^{\mp}$, we have performed a parameter space scan consistent with theoretical constraints as well as current collider, flavor, and EWPO data. 

We have shown that $\gamma\gamma$ collisions can rival the discovery potential of the conventional $e^+e^-$ mode for probing doubly charged Higgs bosons through the $4\ell+\slashed{E}_T$ signature. Despite the reduced effective luminosity resulting from the photon spectrum, the significantly enhanced production $\gamma\gamma$ cross sections, exceeding those in electron-positron collisions by more than one order of magnitude, compensate for this limitation. 

A detector-level signal-to-background analysis, using an ILC detector card for illustration, shows that a discovery significance at the $5\sigma$ level can be achieved for viable BPs at $\sqrt{s}=830$ and $1245$ GeV for the three luminosity configurations considered, $50$, $100$, and $150$ fb$^{-1}$. These findings show that photon colliders can provide a competitive and complementary platform for exploring doubly charged scalar states, (and, collaterally, also singly ones) in extended Higgs sectors.
		
		\vspace{6pt}
		\label{conlusion}	
		\section*{ACKNOWLEDGMENTS}
		A. Arhrib is supported by the Arab Fund for economic and social development. M. Boukidi acknowledges the support of the Narodowe Centrum Nauki under OPUS Grant No. 2023/49/B/ST2/03862 as well as the use of the PALMA II high-performance computing cluster at the University of M\"unster, subsidized by the DFG (INST 211/667-1). K. Goure would like to thank CNRST/HPC-MARWAN for technical support. S. Moretti is supported in part through the NExT Institute and  the STFC CG ST/X000583/1.
		
		\bibliographystyle{JHEP}
		\bibliography{bibliography}

	\end{document}